\documentclass[amsmath,amssymb,aps]{revtex4-2}

\usepackage{graphicx}
\usepackage{dcolumn}
\usepackage{bm}

\usepackage{comment}
\usepackage{braket}
\usepackage{bbm}
\usepackage{bm}
\usepackage{tikz}
\usetikzlibrary{shapes.geometric}
\usepackage{color}
\usepackage{graphicx}
\usepackage{enumerate}
\usepackage{hyperref}

\newcommand{\be}{\begin{equation}}
\newcommand{\ee}{\end{equation}}
\newcommand{\ba}{\begin{aligned}}
\newcommand{\ea}{\end{aligned}}

\newcommand{\bs}[1]{\bm{#1}}

\newcommand{\pf}{\mathrm{pf} }

\newcommand{\tr}{\mathrm{tr}}

\begin{document}


\title{Universality in the tripartite information after global quenches:\\spin flip and semilocal charges}

\author{Vanja Mari\'c}
 \affiliation{Universit\'e Paris-Saclay, CNRS, LPTMS, 91405, Orsay, France.}

\begin{abstract}
We study stationary states emerging after global quenches in which the time evolution is under local Hamiltonians that possess semilocal conserved operators. In particular, we study a model that is dual to quantum XY chain. We show that a localized perturbation in the initial state can turn an exponential decay of spatial correlations in the stationary state into an algebraic decay. We investigate the consequences on the behavior of the (R\'enyi-$\alpha$) entanglement entropies, focusing on the tripartite information of three adjacent subsystems. In the limit of large subsystems, we show that in the stationary state with the algebraic decay of correlations the tripartite information exhibits a non-zero value with a universal dependency on the cross ratio, while it vanishes in the stationary state with the exponential decay of correlations.

\end{abstract}

\maketitle


\section{Introduction}

This is the third paper in a series of works about the tripartite information in the stationary states after global quenches, the first two being ref. \cite{Maric2022Universality} and ref. \cite{Maric2023Universality}, that will be in the following referred to as Paper I and Paper II respectively. Paper II proves the results announced in Paper I concerning quantum quenches from ground states of critical Hamiltonians and bipartitioning protocols. The present paper completes the proofs of the results announced in Paper I and deals with quenches in which semilocal conservation laws affect the dynamics.

Quantum quench is a protocol in which the system is prepared in the ground state of some local Hamiltonian and is suddenly let to evolve with a different local Hamiltonian. It is perhaps the simplest way to induce non-equilibrium dynamics and as such it has been thoroughly investigated~\cite{Polkovnikov2011Colloquium,Eisert2015Quantum,Gogolin2016Equilibration}. In global quenches, in particular, the two Hamiltonians are macroscopically different. It has been established that in global quenches quite general isolated quantum many-body systems locally relax to thermal states, as explained by the eigenstate thermalization hypothesis~\cite{Rigol2008,deutsch91,Srednicki1994Chaos, Deutsch_2018}. Integrable systems are an exception \cite{Kinoshita2006}. Their stationary properties are instead captured by generalized Gibbs ensembles (GGE) \cite{Rigol2007Relaxation,Doyon2017Thermalization,Vidmar2016Generalized,Essler2016Quench}, which carry memory of additional conserved charges.

While most of the established results on global quenches concern translationally invariant systems, there has been also a lot of effort to relax the assumption of translational invariance. Bipartitioning protocols \cite{Alba2021Generalized} have been studied substantially, in which the initial state consists of two macroscopically different parts, for example a domain wall. The theory of generalized hydrodynamics~\cite{Bertini2016Transport,Castro-Alvaredo2016Emergent,Bertini2016Determination} (GHD) explains that in bipartitioning protocols the system can at late times be described in terms of locally-quasi-stationary states, that depend on the ratio of the distance from the inhomogeneity and the time. At infinite time (in the thermodynamic limit) around the initial inhomogeneity, in particular, the system relaxes to a stationary state, usually called non-equilibrium stationary state (NESS). 

One of the standard tools to capture the universal properties of systems are the entanglement entropies. At criticality the entanglement entropy of a connected block exhibits a simple logarithmic scaling with the subsystem size, where the prefactor is proportional to the central charge of the underlying conformal field theory (CFT) \cite{Holzhey1994Geometric,Calabrese2004Entanglement,Korepin2004PRL,Vidal2003,Calabrese2009Entanglement}. The entanglement entropies have been thoroughly studied also after global quenches and it has been established that they saturate to a value that is extensive with the size of the subsystem \cite{Calabrese2005Evolution,Sotiriadis2008,Bastianello2018Spreading,Bertini2022Growth,Alba2017Entanglement,Alba2018Entanglement,Casini2016spread,Liu2014,Skinner2019}. In Papers I and II it has been pointed out that the entanglement entropy of a connected block after a global quench can exhibit, both in translationally invariant quench protocols and in NESS, besides the extensive term, also a subdominant term that grows logarithmically with the subsystem size (see eq. \eqref{single block entropy scaling}), similarly to the universal leading term in CFT. Such subextensive logarithmic terms have been found also in other works on NESS \cite{Eisler2014,Fraenkel2021,FagottiMaricZadnik2022}.

Because of the similarity of the subleading logarithmic term to the universal leading term in CFT, it is desirable to consider a quantity that removes the extensive and the boundary contributions of the entropies, leaving a potentially universal quantity. In Papers I and II a special linear combination of the entropies with such a property, namely the tripartite information \cite{Cerf1998Information} of three adjacent blocks, has been considered for a wide class of quench protocols in non-interacting spin chains. It has been shown that quenches from critical points and bipartitioning protocols can exhibit stationary states with a non-zero tripartite information, which is accompanied by the subdominant logarithmic term in the entanglement entropy and the existence of spatial correlation functions that decay algebraically, similarly to CFT. In 1+1-dimensional CFT the tripartite information of three adjacent blocks is a model-dependent function with a universal dependency on the cross ratio \cite{Calabrese2009Entanglement,Calabrese2009Entanglement1}. {We note that the term ``universal" here does not mean the independence of the parameters of the model, but refers to the property that the tripartite information of three adjacent blocks depends on their lengths only through the cross ratio and is independent of the further details of the configuration. On the lattice this universal property is achieved for \textit{large} subsystem lengths.} Remarkably, tripartite information in the stationary states of the aforementioned global quench protocols {also exhibits this universality.} A property making it different from CFT is the existence of a nonzero ``residual tripartite information", introduced in Paper I. 

Localized perturbations of initial states are normally not expected to have any macroscopic effects at large times. For systems of non-interacting fermions it has been proven under quite general conditions that the equilibration towards a GGE is resilient to localized perturbations, given that the initial state has a finite correlation length and that the evolution Hamiltonian is translationally invariant \cite{Gluza2019Equilibration}. However, memory effects can arise following a \textit{local} quench in quantum spin chains, related to the non-locality of the mapping between spins and fermions \cite{Zauner2015,eisler2020Front,Eisler2016Universal,Gruber2021Entanglement,Eisler2018Hydrodynamical} and to jammed states \cite{Bidzhiev2022Macroscopic,Zadnik2022Measurement}. Recently it has been found that the system can keep the memory of localized perturbations even following a \textit{global} quench, if it possesses semilocal conserved operators \cite{Fagotti2022Global}. In such a system a single spin flip in the initial product state can induce a subextensive logarithmic term in the entropy after a global quench \cite{FagottiMaricZadnik2022}, which hints at a possible nontrivial tripartite information.

Semilocal charges are conserved operators whose density does not necessarily commute with distant localized operators. Sometimes their densities can be interpreted as semi-inifinte strings. Semilocal charges enable the existence of string order after global quenches and, in general, they have to be included in the GGE to capture correctly the properties of the stationary state \cite{FagottiMaricZadnik2022}. Furthermore, in such systems a single localized perturbation in the initial state can change the magnetization in the stationary state \cite{Fagotti2022Global} and induce time growth of macroscopic entanglement \cite{Bocini2023growing}. Similar semilocal symmetries have also been discussed in the context of bistability of driven-dissipative fermionic systems \cite{Alaeian2022}.

Here we study a system with semilocal charges and show that in such a system a single spin flip in the initial product state can turn an exponential decay of spatial correlations in the stationary state into an algebraic decay. {More precisely, we show that, while the stationary state with an exponential decay of spatial correlations emerges in the quench from a translationally invariant initial product state, flipping a single spin in the initial state results in NESS (corresponding to the ray $d/t=0$, where $d$ is the distance from the initial inhomogeneity and $t$ is time) with an algebraic decay of spatial correlations. Even though it is possible that the two quench protocols exhibit different time scales on which the stationary states are reached, we stress that the inifinite time limit is well defined in both cases.} Then we proceed to studying the consequences on the behavior of the entanglement entropies, focusing on the tripartite information. We show that {NESS arising from the state with} a single spin flip {displays the} behavior of the tripartite information found in bipartitioning protocols and quenches from ground states of critical systems.

The paper is organized as follows. In the remainder of the introduction we discuss the tripartite information (section \ref{section tripartite information}) and we introduce the model and the quench protocols under consideration (section \ref{section model and quench protocols}). In section \ref{section results} we present the results of the paper. Their derivation is presented afterwards, in section \ref{section methods}. Conclusions are drawn in section \ref{section conclusions}.

\subsection{Tripartite Information}\label{section tripartite information}

\begin{figure}

\begin{tikzpicture}[scale=0.4]
     \draw[black,line width=0.6pt] (-1,0) to (27,0);
    \foreach \x in {0,...,26}
    \filldraw[ball color=blue!20!white,opacity=0.75,shading=ball] (\x,0) circle (6pt);

    \node[anchor=south] at (6.5,0.6) {\large $A$};
     \node[anchor=south] at (13,0.6) {\large $B$};
      \node[anchor=south] at (19.5,0.6) {\large $C$};
    \draw[fill=blue,opacity=0.2, rounded corners = 2,thick] (3.6,-0.5) rectangle ++(5.8,1);
     \draw[fill=blue,opacity=0.2, rounded corners = 2,thick] (9.6,-0.5) rectangle ++(6.8,1);
 \draw[fill=blue,opacity=0.2, rounded corners = 2,thick] (16.6,-0.5) rectangle ++(5.8,1);
    \end{tikzpicture}
    \caption{We compute the tripartite information of adjacent blocks $A$, $B$ and $C$ in stationary states after global quenches.}
    \label{fig disjoint blocks}
\end{figure}
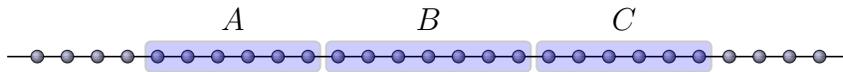

Given a subsystem $A$ described by a reduced density matrix $\rho_A$, the von Neumman entanglement entropy is defined as
\begin{equation}
    S_1(A)=-\tr \left(\rho_A\log \rho_A\right).
\end{equation}
It corresponds to the limit $\alpha\to 1$ of the R\'enyi entanglement entropies
\begin{equation}
    S_\alpha(A)=\frac{1}{1-\alpha}\log\tr\left(\rho_A^\alpha\right)  .
\end{equation}
Typically, R\'enyi entropies for $\alpha=2,3,\ldots$ are more accessible to computations than the von Neumann entropy and sometimes the latter can be obtained from the former using the replica trick~\cite{Calabrese2004Entanglement}.

In pure states the entanglement entropies are a measure of quantum entanglement. In mixed states, such as thermal states and GGEs, this is not anymore the case, as the entropies carry contributions that are extensive with the size of the subsystem and that are largely due to classical correlations. Such extensive contributions are cancelled in the (R\'enyi-$\alpha$) mutual information. Given two subsystems, $A$ and $C$, the mutual information \cite{Cerf1998Information} is defined as
\begin{equation}
I_2^{(\alpha)}(A,C)=S_\alpha(A)+S_\alpha(C)-S_\alpha(AC) \;.
\end{equation}
Here and in the rest of the paper $AC$ stands for the union $A\cup C$ of two sets $A,C$. For large disjoint blocks $A,C$ in the configuration of figure \ref{fig disjoint blocks} mutual information cancels both the extensive and the boundary contributions of the entropies. Mutual information for $\alpha=1$ is non-negative and is a measure of total correlations, classical and quantum, between $A$ and $C$ \cite{Groisman2005}. Moreover, mutual information provides an upper bound for connected correlation functions \cite{Wolf2008}. 
We mention that when $A$ and $C$ constitute the whole lattice, one being the complement of the other, there are area laws~\cite{Wolf2008,Kuwahara2021,Eisert2010Area_laws,Bernigau2015,Lemm2022,Alhambra2022} that render mutual information finite in thermal states.

Another quantity that cancels the extensive and the boundary contributions of the entropies is the (R\'enyi $\alpha$) tripartite information \cite{Cerf1998Information}, defined for three subsystems $A,B,C$ as
\begin{equation}\label{eq:tripartite definition in terms of S}
    I_3^{(\alpha)}(A,B,C)= S_\alpha(A)+S_\alpha(B)+S_\alpha(C)-S_\alpha(AB)-S_\alpha(A C)-S_\alpha(B C)+S_\alpha(ABC) \; ,
\end{equation}
which can be expressed in terms of the mutual information as
\begin{equation}
    I_3^{(\alpha)}(A,B,C)=I_2^{(\alpha)}(A,B)+I_2^{(\alpha)}(A,C)-I_2^{(\alpha)}(A,BC) \; .
\end{equation}
For three adjacent blocks, presented in figure \ref{fig disjoint blocks}, unlike the mutual information, it has a desirable property of remaining bounded in the limit $1\ll |B|\ll |A|,|C|$, that will be of central interest. Here and in the following $|A|$ stands for the size of $A$. Namely, let us suppose that the entanglement entropy of a (large) connected block scales as
\begin{equation}\label{single block entropy scaling}
    S_\alpha(A)=a_\alpha |A|+b_\alpha \log|A|+c_\alpha \; ,
\end{equation}
for some constants $a_\alpha,b_\alpha,c_\alpha$. This is the case both in CFT ($a_\alpha=0, b_\alpha\neq 0$) and in the stationary states studied in this work ($a_\alpha\neq 0$). Then, for the configuration in figure \ref{fig disjoint blocks} the relation between the tripartite information of $A,B,C$ and the mutual information of $A,C$ is (in the limit of large blocks)
\begin{equation}
    I_2^{(\alpha)}(A,C)=-b_\alpha\log(1-x)+I_3^{(\alpha)}(A,B,C) \; ,
\end{equation}
where
\begin{equation}\label{eq:cross ratio definition}
    x=\frac{|A||C|}{(|A|+|B|)(|B|+|C|)}
\end{equation}
is the cross ratio. The mutual and the tripartite information can thus be simply obtained one from another, and the tripartite information describes a subleading bounded contribution to the mutual information in the limit $x\to 1^-$, that corresponds to the limit $1\ll |B|\ll |A|,|C|$. Note that, in general, the tripartite information can have any sign \cite{Casini2009Remarks,Fagotti2012New} (even for $\alpha=1$).

The tripartite information has been studied in many settings. In the context of topological order in two space dimensions it is usually called simply ``topological entanglement entropy" \cite{Kitaev2006Topological}. It has been studied in quantum field theory, both in 1+1 dimensions \cite{Casini2009Remarks,Caraglio2008Entanglement,Furukawa2009Mutual,Rajabpour2012,Calabrese2009Entanglement1,Calabrese2011Entanglement,Coser2014OnRenyi,Ruggiero2018,Alba2010Entanglement,Blanco2011,Alba2011Entanglement,Fagotti2010disjoint,Fagotti2012New,Fries2019,Balasubramanian2011,Grava2021,Ares2021Crossing}, mainly CFT, and in higher dimensions \cite{Casini2009Remarks,Agon2016,Agon2022,AliAkbari2021}. In holographic theories it has been shown that the tripartite information is never positive \cite{Hayden2013Holographic}. The tripartite information has also been studied in continuously monitored chains \cite{Carollo2022}, on Hamming graphs \cite{Parez2022Multipartite} and, with the partition at different time slices, as a diagnostic tool for quantum scrambling \cite{Hosur2016Chaos,Schnaack2019Tripartite,Sunderhauf2019Quantum,Kuno2022}. Finally, it has also been addressed after global quenches in closed quantum systems \cite{Maric2022Universality,Maric2023Universality,Parez2022,Caceffo2023}.

In this work we focus on the tripartite information of three large adjacent subsystems embedded in an infinite chain (see figure \ref{fig disjoint blocks}). We refer the reader to section 2 of paper II for a discussion of the behavior of the tripartite information in this setting for different systems. Here we note that it is expected to vanish when the spin correlation functions decay exponentially with distance, as is the case in thermal states \cite{Bluhm2022exponentialdecayof} and ground states of gapped Hamiltonians \cite{Hastings2006}. An exception are ground states of conformal critical systems, where the tripartite information is a model dependent function of the cross ratio in \eqref{eq:cross ratio definition} \cite{Calabrese2009Entanglement1,Calabrese2009Entanglement}.
For example, in the ground state of the XX chain the R\'enyi-$2$ tripartite information reads \cite{Furukawa2009Mutual,Calabrese2009Entanglement1,Fagotti2012New}
\begin{equation}\label{tripartite XX ground state alpha 2}
    I_3^{(2)}(A,B,C)=-\log 2 +\log\left(1+\sqrt{1-x}+\sqrt{x}\right) \;.
\end{equation}

Papers I and II identified different global quench protocols that result in a stationary state in which the spin correlation functions decay algebraically (in space) and allow for non-zero tripartite information: bipartitioning protocols and quenches from ground states of critical systems. In the stationary states of these protocols the entropy of a large single (connected) block $A$ satisfies scaling \eqref{single block entropy scaling} with both $a_\alpha$ and $b_\alpha$ nonzero. The tripartite information of three large adjacent blocks in these stationary states exhibits a universal dependency on the cross ratio \eqref{eq:cross ratio definition}, similarly to critical systems. For example, the quench from the domain wall state $\ket{\ldots \uparrow\uparrow\uparrow\downarrow\downarrow\downarrow\ldots}$ under the XY chain Hamiltonian results in the stationary state with the R\'enyi-$2$ tripartite information
\begin{equation}\label{tripartite domain wall alpha 2}
    I_3^{(2)}(A,B,C)=-\log 2 +\log\left(1+\sqrt{1-x}\right) \;,
\end{equation}
which corresponds to \eqref{tripartite XX ground state alpha 2} with the last term in the logarithm dropped.

Differently from CFT, the tripartite information in the aforementioned stationary states is not symmetric under the interchange $x\leftrightarrow 1-x$. This property is particularly transparent in the limit $x=1^-$, which corresponds to the limit in which the central subsystem is much smaller than the others, i.e. the limit $1\ll |B|\ll |A|,|C|$. In this limit the tripartite information has a nonzero value, termed ``residual tripartite information" in Paper I. The nonzero value is $-\log 2$ and it is common to all studied quench protocol with non-zero tripartite information. Moreover, this value is independent of the R\'enyi index $\alpha$, so it applies also to the von Neumann tripartite information. It should be contrasted to the zero value found in equilibrium at any temperature, irrespectively of criticality, or in other non-equilibrium settings, such as after quenches from ground states of gapped Hamiltonians. We note that there are also trivial examples with nonzero residual tripartite information, such as the GHZ state $\ket{\textrm{GHZ}}=(\ket{\ldots\uparrow\uparrow\uparrow\ldots}+\ket{\ldots\downarrow\downarrow\downarrow\ldots})/\sqrt{2}$, which has tripartite information $+\log 2$ independently of the size of the subsystems (as long as the subsystems do not comprise the whole system, in which case the tripartite information would be zero since the state is pure), but such examples differ from the systems we are interested in because they violate clustering.

We note that algebraic decay of some spin correlation functions is not a sufficient condition for a non-zero tripartite information, as illustrated by the global quench from the ground state of the Ising chain with a critical transverse field (see Papers I and II). The latter reaches a stationary state in which some spin correlation functions decay with distance $r$ as $1/r^4$, yet the tripartite information is zero. This phenomenology seems to be related to the degree in the power law, since quenches with non-zero tripartite information posses correlations that decay only as $1/r^2$.

\subsection{Model and Quench Protocols}\label{section model and quench protocols}

\paragraph{Model} We study the time evolution governed by the dual XY chain \cite{Zadnik2021The,Fagotti2022Global}, given by the Hamiltonian
\begin{align}
    \bs H = \sum_{\ell=-\infty}^\infty\bs\sigma^x_{\ell-1}(J_x\bs I-J_y\bs\sigma^z_\ell)\bs\sigma^x_{\ell+1} \label{eq:dual_xy} \; ,
\end{align}
where we assume $0<|J_y|<|J_x|$. A particularly interesting property of the model is the existence of semilocal charges \cite{Fagotti2022Global,FagottiMaricZadnik2022}, introduced in the following. We also mention that the special point $J_x=J_y$ has been studied for its kinetic constraints \cite{Zadnik2023slow,Pai2020,Eck2023xxz}.

The Hamiltonian is invariant under spin flip $\mathcal{P}_\sigma^z$,
where
\begin{equation}
   \mathcal{P}_\sigma^z[\bs O]=\lim_{n\to\infty} \left(\prod_{\ell=-n}^n \bs \sigma_{\ell}^z \right) \bs O \left( \prod_{\ell'=-n}^n \bs \sigma_{\ell'}^z\right)
\end{equation}
for some localized operator $\bs O$. Here and in the following we say that a local operator $\bs Q=\sum_{\ell\in \mathbb Z}\bs q_\ell$, where the density $\bs q_\ell$ is localized (meaning finite support), is even/odd under some transformation $\mathcal{P}$ if $\mathcal{P}[ \bs q_\ell]=\pm \bs q_\ell$ respectively.
We will use the bold notation exclusively for operators defined on the whole, infinite, chain.

A charge $\bs Q$ is an operator commuting with the Hamiltonian ($ [\bs Q, \bs H]=0 $). While for local charges $\bs Q=\sum_{\ell\in \mathbb Z}\bs q_\ell$ the density $\bs q_\ell$ is localized around site $\ell$, semilocal charges of this model are characterised by density $\bs q_\ell$ with support on all sites on one side of $\ell$. Namely, defining the operators $\bs\Pi^z_{\sigma,+}(\ell)$ by
\begin{equation}\label{eq:Pi z semilocal operator}
    \bs \Pi^z_{\sigma,+}(\ell)\bs \sigma_j^{x,y} =\begin{cases}
        \bs \sigma_j^{x,y} \bs \Pi^z_{\sigma,+}(\ell) & j<\ell \\
         - \bs \sigma_j^{x,y} \bs \Pi^z_{\sigma,+}(\ell) & j\geq\ell   \end{cases}
         , \qquad  \left[\bs \Pi^z_{\sigma,+}(\ell), \bs\sigma_j^z\right]=0, \qquad \left(\bs \Pi^z_{\sigma,+}(\ell)\right)^2= \bs 1 ,
\end{equation}
which can be thought of as semi-infinite strings $\bs\sigma_\ell^z\bs\sigma_{\ell+1}^z\bs\sigma_{\ell+2}^z\ldots$,
an example of a semilocal charge of the model in \eqref{eq:dual_xy} is
\begin{equation}\label{eq:charge 0 - sigma}
    \bs Q^{(0,-)}=\frac{1}{2}\sum_{\ell=-\infty}^\infty \bs\sigma_{\ell-1}^x\left(\bs I-\bs\sigma_\ell^z \right)\bs\sigma_{\ell+1}^y\bs\Pi_{\sigma,+}^z(\ell+2) \; .
\end{equation}
A complete set of one-site shift invariant semilocal charges of the model is given in section \ref{section semilocal charges}. Their construction is directly related to the Kramers-Wannier duality. Semilocal charges are particularly interesting because they allow the sytem to keep the memory of localized perturbations in the initial state. Namely, while localized perturbations can affect the expectation value of localized charge densities $\bs q_\ell$ at most for several sites $\ell$, the effect on semilocal charges can be drastic.

\paragraph{Quench protocols} In this work we study and compare two different global quench protocols:
\begin{enumerate}[{1)}]
\item \label{quench protocol 1}\textit{all-spin-up} quench protocol: time evolution of a translationally invariant product state, $\ket{\Psi(t)}=e^{-i\bs H t}\ket{\Uparrow}$;
\item \label{quench protocol 2} \textit{flipped-spin} quench protocol: time evolution of the state in \ref{quench protocol 1}) with a flipped spin at a single site, $\ket{\Psi(t)}=e^{-i\bs H t}\bs \sigma_0^x\ket{\Uparrow}$.
\end{enumerate}
Here and in the rest of the paper by $\Uparrow$ we denote an inifinite string of $\uparrow$. We can thus also write $\bs \sigma_0^x\ket{\Uparrow}=\ket{\Uparrow \downarrow_0 \Uparrow}$. In the flipped-spin protocol the system will for large times reach quasi-stationary states at different rays $\zeta=d/t$, where $d$ is the distance from the site with the flipped spin. When speaking about the stationary state we will always refer to the NESS reached at infinite time around the initial inhomogeneity, corresponding to the ray $\zeta=0$.

\paragraph{Kramers-Wannier transformation} We consider a transformation that differs  from the standard Kramers-Wannier duality map, responsible for the self-duality of the transverse-field Ising model, just in an additional rotation,
\be\label{eq:dualTDeven}
\bs\tau_j^x=\bs\sigma_{j-1}^x\bs\sigma_{j}^x\, ,\qquad
\bs \tau_j^z\bs \tau_{j+1}^z=\bs\sigma_j^z\, ,
\ee
that we will also refer to as Kramers-Wannier transformation. We will refer to the representations of the theory in terms of the $\bs \tau$ operators as the dual picture.

The transformation preserves the algebra of Pauli matrices, i.e. the $\bs \tau$ operators satisfy the same algebra as the $\bs \sigma$ ones. Relation \eqref{eq:dualTDeven} specifies the transformation only for localized operators even under $\mathcal{P}_\sigma^z$, which are transformed to operators that are localized also in the dual picture and that are, moreover, even under $\mathcal{P}^x_\tau$, where
\begin{equation}
   \mathcal{P}^x_\tau[\bs O]=\lim_{n\to\infty} \left(\prod_{\ell=-n}^n \bs \tau_{\ell}^x \right) \bs O \left( \prod_{\ell'=-n}^n \bs \tau_{\ell'}^x\right) \; .
\end{equation}
In this work, the full mapping is needed only for the discussion of semilocal charges. Details are given in section \ref{section Kramers-Wannier}.

Kramers-Wannier transformation maps the dual XY chain \eqref{eq:dual_xy} to the XY chain
\begin{equation}
 \bs H=\sum_{\ell=-\infty}^\infty\left(J_x\bs\tau_\ell^x\bs\tau_{\ell+1}^x +J_y\bs\tau_\ell^y\bs\tau_{\ell+1}^y \right)\;,\label{eq:XY_model}
\end{equation}
a very well known model mappable to free fermions~\cite{Lieb1961}, with substantially developed techniques for quantum quenches \cite{Calabrese2011Quantum,Calabrese2012Quantum,Calabrese2012QuantumII,Fagotti2013Reduced,Fagotti2020}. The state of all spin up maps into itself, while the state with a spin flip maps into a domain wall state. Namely, we have the identification $\ket{\Uparrow}^{(\sigma)}=\ket{\Uparrow}^{(\tau)}$ and $\ket{\Uparrow\downarrow_0 \Uparrow}^{(\sigma)}=\ket{\Uparrow\Downarrow}^{(\tau)}$ with the domain wall between sites $0$ and $1$. Note that these are not unique identifications and one could, for example, identify $\ket{\Uparrow}^{(\sigma)}$ with $\ket{\Downarrow}^{(\tau)}$ instead of $\ket{\Uparrow}^{(\tau)}$ or with any linear combination of the two. In a finite system \cite{Fagotti2022Global} there is a unique choice, {related to the boundary conditions at the edge of the system. Since we let the boundaries to infinity, for our purposes} different choices are equivalent. In any case, a single spin flip changes the state macroscopically in the dual picture. {We note that these aspects of the mapping are very similar to the hidden symmetry-breaking picture of the symmetry-protected topological phases \cite{Kennedy1992Hidden,Else2013Hidden,Duivenvoorden2013From}.}

\paragraph{Reduced density matrices and tripartite information}

We stress it is a highly non-trivial question what are the effects of the Kramers-Wannier transformation on the tripartite information. For example, the Jordan-Wigner transformation is also a duality transformation and there are important differences between the entanglement entropy of disjoint blocks of spins and fermions \cite{Fagotti2010disjoint,Igloi2010}, and therefore also in the tripartite information. In fact, this difference is crucial for the phenomenology discovered in Papers I and II. However, it turns out that the Kramers-Wannier transformation in the studied stationary states does not affect the tripartite information of three large adjacent blocks.

For a system in a state $\ket{\Psi}$, the reduced density matrix for subsystem $X$ is given by
\begin{equation}\label{density matrix sigma basis tensor}
        \rho_X=\frac{1}{2^{|X|}}\sum_{\gamma_{\ell}\in\{0,x,y,z\},\ell\in X} \bra{\Psi}\prod_{\ell\in X}\bs \sigma^{\gamma_\ell}_\ell \ket{\Psi}\bigotimes_{\ell\in X} \sigma^{\gamma_\ell} \;.
\end{equation}
Here the sum is over an orthogonal basis of operators on $X$, given by all possible products of Pauli matrices, and we use the standard convention $\sigma^0\equiv \mathbb{I}$. While the sites $\ell$ of some physical subsystem $X$ are associated to the indices of the operators $\bs\sigma_{\ell}^\gamma$, in the dual picture we associate the notion of subsystem to the indices of the $\bs\tau^\gamma_\ell$ operators. Accordingly, in the dual picture it is natural to consider the density matrix
\begin{equation}\label{density matrix tau basis tensor}
    \rho_X^\tau=\frac{1}{2^{|X|}}\sum_{\gamma_{\ell}\in\{0,x,y,z\},\ell\in X} \bra{\Psi}\prod_{\ell\in X}\bs \tau^{\gamma_\ell}_\ell \ket{\Psi}\bigotimes_{\ell\in X} \sigma^{\gamma_\ell} \; .
\end{equation}
For the studied quench protocols the density matrix in \eqref{density matrix tau basis tensor} can be assessed using the techniques developed for the model in \eqref{eq:XY_model}. However, it is a non-trivial question how are density matrices \eqref{density matrix sigma basis tensor} and \eqref{density matrix tau basis tensor} related, or how are the corresponding entanglement entropies related. These questions are some of the main problems we tackle with in this work. As already mentioned, for a single block of spins $X$ this problem has been addressed in \cite{FagottiMaricZadnik2022}. The result is that density matrices \eqref{density matrix sigma basis tensor} and \eqref{density matrix tau basis tensor} give different entanglement entropies in general, but the difference is bounded by a constant independent of the subsystem size $|X|$. However, from this result we cannot conclude about the tripartite information \eqref{eq:tripartite definition in terms of S}, which has also a contribution from the entropy of disjoint blocks and which is, moreover, bounded itself. Thus, technically, the main problem of this work is to study the density matrix \eqref{density matrix tau basis tensor} when $X$ consists of disjoint blocks $A,C$, as presented in section \ref{section methods}. In the end we are able to conclude that in the studied stationary states the differences in the entropies corresponding to density matrices \eqref{density matrix sigma basis tensor} and \eqref{density matrix tau basis tensor} cancel in the tripartite information of three large adjacent blocks.

\section{Results}\label{section results}

\begin{figure}
    \centering
    \hspace{-1.5 cm}
    \includegraphics[width=0.7\textwidth]{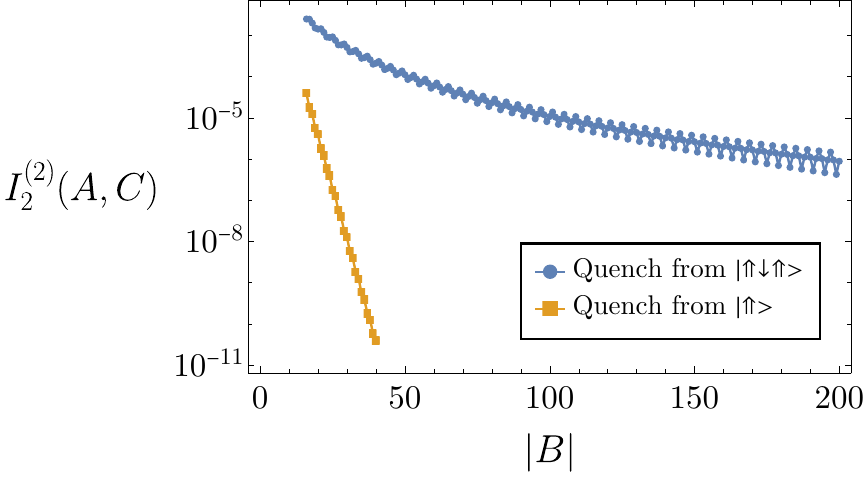}
    \caption{R\'enyi-$2$ mutual information between subsystems $A$ and $C$ (see figure \ref{fig disjoint blocks}) for $|A|=|C|=16$ and varying distance $|B|$, in the stationary states following quench protocols \ref{quench protocol 1} and \ref{quench protocol 2}. In the stationary state of protocol \ref{quench protocol 1} mutual information decays exponentially with distance ($y$-axis is in the log scale), while in the stationary state of protocol \ref{quench protocol 2} it decays algebraically.}
    \label{fig:MutualInformation}
\end{figure}

\subsection{Correlation functions}\label{section results spin correlations}
Computing the spin-correlation functions in the stationary states of protocols \ref{quench protocol 1} and \ref{quench protocol 2} is a straightforward application of the Kramers-Wannier duality and the Jordan-Wigner transformation for the dual model, as commented in section \ref{section jordan-wigner transformation}.

The connected correlation functions of the $z$-components of spins in the stationary state following quench protocol \ref{quench protocol 1} are exactly zero,
\begin{equation}
   \textrm{protocol \ref{quench protocol 1}:}\qquad \braket{\bs\sigma_0^z\bs\sigma_r^z}-\braket{\bs\sigma_0^z}\braket{\bs\sigma_r^z}=0, \qquad r\geq 1,
\end{equation}
while in the stationary state of protocol \ref{quench protocol 2} they exhibit algebraic decay with distance,
\begin{equation}\label{connected Z spin correlation functions protocol 2}
    \textrm{protocol \ref{quench protocol 2}:}\qquad\braket{\bs\sigma_0^z\bs\sigma_r^z}-\braket{\bs\sigma_0^z}\braket{\bs\sigma_r^z}\simeq\frac{16}{\pi^4}\frac{1}{r^4}, 
\end{equation}
where $\simeq $ means asymptotically equal.
The connected spin-correlation functions of other spin components are exactly zero (for large enough distance) both in the stationary state following quench protocol \ref{quench protocol 1} and the one following protocol \ref{quench protocol 2},
\begin{equation}
    \textrm{protocols \ref{quench protocol 1} and \ref{quench protocol 2}:}\qquad\braket{\bs\sigma_0^\gamma\bs\sigma_r^\gamma}-\braket{\bs\sigma_0^\gamma}\braket{\bs\sigma_r^\gamma}=0, \qquad \gamma=x,y, \ r\geq 5. 
\end{equation} 

These results deal only with the correlation functions of operators with support on one site, which already establish that there are algebraically decaying correlations in the stationary state of quench protocol \ref{quench protocol 2}. For the stationary state of quench protocol \ref{quench protocol 1} we provide evidence that there are no connected correlation functions decaying algebraically by computing the R\'enyi-$2$ mutual information. The results are presented in figure \ref{fig:MutualInformation}. In the stationary state of protocol \ref{quench protocol 1} the mutual information $ I_2^{(2)}(A,C)$ decays exponentially with the distance between $A$ and $C$, while it decays only algebraically in the stationary state of protocol \ref{quench protocol 2}. Assuming that the von Neumann mutual information $I_2^{(1)}(A,C)$ follows an analogous behavior, we can conclude that all connected correlation functions decay exponentially in protocol \ref{quench protocol 1} \cite{Wolf2008}.

We note that in the stationary state of quench protocol \ref{quench protocol 2} it is possible to construct operators whose connected correlation functions decay only as $1/r^2$, as opposed to the $1/r^4$ decay of the spin correlation functions in eq. \eqref{connected Z spin correlation functions protocol 2}. This is the case for the operator $\bs\xi_\ell\equiv \bs\sigma_{\ell-1}^x\bs\sigma_{\ell}^z\bs\sigma_{\ell+1}^x$, with the correlation functions
\begin{equation}\label{connected more complicated correlation functions protocol 2}
\begin{split}
    \textrm{protocol \ref{quench protocol 2}:}\qquad \braket{\bs\xi_{0}\bs\xi_{2r-1}}-\braket{\bs\xi_0}\braket{\bs\xi_{2r-1}}\simeq-\frac{4}{\pi^2r^2} ,\qquad \braket{\bs\xi_{0}\bs\xi_{2r}}-\braket{\bs\xi_0}\braket{\bs\xi_{2r}}=0, \; r\geq 1.
    \end{split}
\end{equation}

\begin{figure}
    \centering
    \includegraphics[width=0.6\textwidth]{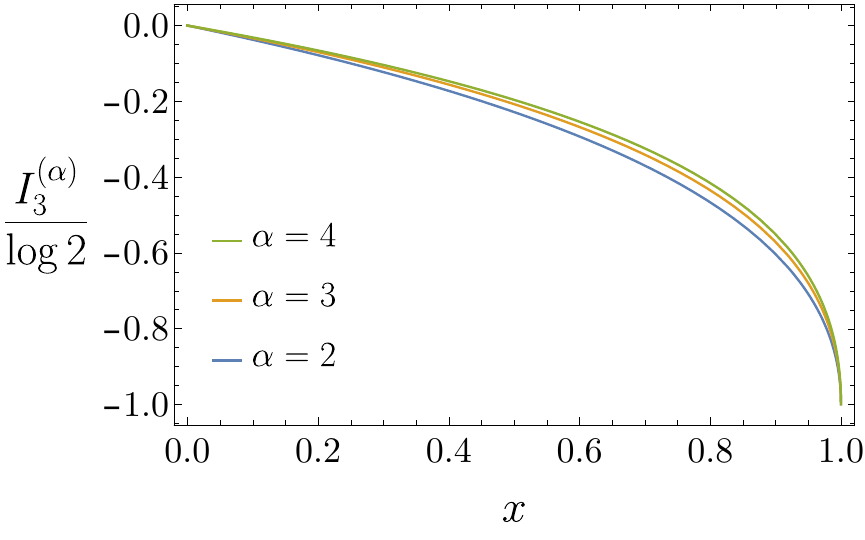}
    \caption{The analytical prediction for the R\'enyi-$\alpha$ tripartite information of three large adjacent blocks (see figure \ref{fig disjoint blocks}) in the stationary state of the flipped-spin quench protocol, given by \eqref{tripartite domain wall any alpha}, as a function of the cross ratio \eqref{eq:cross ratio definition}, for different values of $\alpha$. The curves do not differ much. The R\'enyi-$\alpha$ tripartite information for any $\alpha$ is equal to zero for $x=0^+$ and equal to $-\log 2$ for $x=1^-$. By replica trick the result applies also to the von Neumann tripartite information. A nonzero value in the limit $x=1^-$ was termed ``residual tripartite information" in ref. \cite{Maric2022Universality}.}
    \label{fig:analytical234}
\end{figure}

\subsection{String order}\label{section results string order}

As discussed in ref. \cite{FagottiMaricZadnik2022} in the context of the all-spin-up quench protocol, string order can survive a global quench. Accordingly, the order in the stationary state was termed ``non-equilibrium symmetry-protected topological order". To be more precise, a string of adjacent $\bs\sigma^z_\ell$ operators has a non-zero expectation value in the stationary state even in the limit of infinite length of the string. This is related to the fact that the expectation value of a string of $\bs\sigma^z$ in the dual picture becomes a two-point correlation function (not the connected one) of localized operators. We note that in quench protocol \ref{quench protocol 2} this string order parameter vanishes. Explicitly, we have
\begin{align}
  &  \textrm{protocol \ref{quench protocol 1}:}\qquad \lim_{r\to\infty}\braket{\prod_{\ell=-r}^r\bs\sigma^z_\ell}=\frac{1}{4}\left(1+\frac{J_y}{J_x}\right)^2   , \\ 
  &  \textrm{protocol \ref{quench protocol 2}:}\qquad \lim_{r\to\infty}\braket{\prod_{\ell=-r}^r\bs\sigma^z_\ell}=0  .
\end{align}

Here we point out that there is a string order also in the stationary state of protocol \ref{quench protocol 2}, but given by a different string order parameter, which, on the other hand, vanishes in protocol \ref{quench protocol 1}. Namely, we have
\begin{align}
  &  \textrm{protocol \ref{quench protocol 1}:}\qquad \lim_{r\to\infty}\braket{\bs\sigma_{-r-3}^x\bs\sigma_{-r-2}^z\bs\sigma_{-r-1}^y\left(\prod_{\ell=-r}^{r}\bs\sigma^z_\ell\right)\bs\sigma^y_{r+1}\bs\sigma_{r+2}^z\bs\sigma_{r+3}^x}=0 \; , \\ 
  &  \textrm{protocol \ref{quench protocol 2}:}\qquad \lim_{r\to\infty}\braket{\bs\sigma_{-r-3}^x\bs\sigma_{-r-2}^z\bs\sigma_{-r-1}^y\left(\prod_{\ell=-r}^{r}\bs\sigma^z_\ell\right)\bs\sigma^y_{r+1}\bs\sigma_{r+2}^z\bs\sigma_{r+3}^x}=-\frac{1}{\pi^2}\left(1+\frac{J_y}{J_x}\right)^2 \; .
\end{align}
{The two string operators differ only at the edges, while in the bulk they are given by the on-site action of the symmetry of the model (as in the symmetry-protected topological order at zero temperature \cite{Else2013Hidden,Pollmann12}).} The computation of the string order parameters is commented in section \ref{section jordan-wigner transformation} and appendix \ref{appendix majorana correlators}.

\subsection{Tripartite information}\label{section results tripartite information}

\begin{figure}
    \centering
    \includegraphics[width=0.6\textwidth]{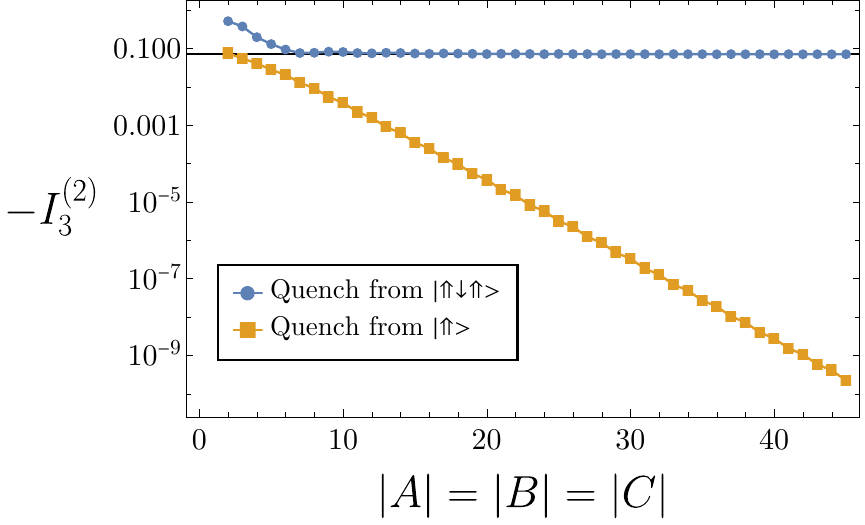}
    \caption{R\'enyi-$2$ tripartite information of three adjacent blocks $A,B,C$ (see figure \ref{fig disjoint blocks}) of equal length, that we vary. In the stationary state of the all-spin-up quench protocol the tripartite information vanishes exponentially with the size of the subsystems, while in the stationary state of the flipped-spin quench protocol it tends to a non-zero value, for which we have analytical prediction, given by eq. \eqref{tripartite domain wall alpha 2} (solid line).}
    \label{fig:FlipAndNoFlip}
\end{figure}

Based on the representation of the reduced density matrix in the dual picture, derived in section \ref{section reduced density matrix in the dual picture}, we argue in section \ref{section invariance} that the tripartite information in the stationary states of the studied global quenches is not affected by the Kramers-Wannier transformation, in the limit of large subsystems. In this way we reduce the problem of computing the tripartite information in the stationary state of quench protocols \ref{quench protocol 1} and \ref{quench protocol 2} to computing the tripartite information in the stationary state of the quench protocol in which the time evolution is with the XY model and the initial state is, respectively, 1) the state of all spin up $\ket{\Uparrow}$, 2) the domain wall state $\ket{\Uparrow\Downarrow}$. These problems have already been studied in details in Papers I and II so we copy the results derived there, and compare them with the exact numerical results for the second R\'enyi entropy, obtained using the methods developed in this work.

In this way we find the following tripartite information in the stationary states of the studied quench protocols, in the limit of large subsystems:
\begin{align}
   & \textrm{protocol \ref{quench protocol 1}:}\qquad I_3^{(\alpha)}(A,B,C)= 0 ,\\
   & \textrm{protocol \ref{quench protocol 2}:}\qquad I_3^{(\alpha)}(A,B,C)=\frac{1}{\alpha-1}\log\Bigl[\sum_{\delta_j\in\{0,\frac{1}{2}\}\atop j=1,\dots,\alpha-1}\Bigl(\tfrac{\Theta(\vec \delta|\hat\tau_x)}{\Theta(\vec 0|\hat\tau_x)}\Bigr)^2\Bigr]-\log 2  \; .  \label{tripartite domain wall any alpha}
\end{align}
Here $\hat \tau_x$ is the $(\alpha-1)\times (\alpha-1)$ period matrix of the Riemann surface $\mathcal R_\alpha$ with elements \cite{Calabrese2009Entanglement1}
\be
[\hat \tau_x]_{\ell n}=\frac{2i}{\alpha}\sum_{k=1}^{\alpha-1}\sin(\tfrac{\pi k}{\alpha})\cos(\tfrac{2\pi k(\ell-n)}{\alpha})\tfrac{P_{(k/\alpha)-1}(2x-1)}{P_{(k/\alpha)-1}(1-2x)}\, ,
\ee
where $P_\mu(z)$ denotes the Legendre functions and $\Theta(\vec z,M)=\sum_{\vec m\in \mathbb Z^{\alpha-1}}e^{i\pi \vec m^t M \vec m+2\pi i \vec m\cdot \vec\delta}$ is the Siegel theta function. The formula \eqref{tripartite domain wall any alpha} is plotted in figure \ref{fig:analytical234} for several values of $\alpha$. For $\alpha=2$ it reduces simply to eq. \eqref{tripartite domain wall alpha 2}. We note that eq. \eqref{tripartite domain wall any alpha} was obtained in Paper I by establishing a correspondence between some contributions to the entanglement entropy of disjoint blocks in the expansion of ref. \cite{Fagotti2010disjoint} and already known CFT resuts~\cite{Coser2016Spin}, while the results for $\alpha=2$ and some simpler approximate formulas for higher $\alpha$ were obtained by a direct computation in Paper II. Note also that the result \eqref{tripartite domain wall any alpha} is independent of the parameters of the model \eqref{eq:dual_xy}.

The tripartite information of three large adjacent blocks is zero in the stationary state of the all-spin-up quench protocol, while in the stationary state of the flipped-spin quench protocol it is nonzero. This conclusion is confirmed by numerical results for the second R\'enyi entropy, presented in figure \ref{fig:FlipAndNoFlip}. There it can be seen that in the stationary state of the all-spin-up quench protocol the tripartite information goes to zero exponentially fast with the subsystem size, while in the stationary state of protocol \ref{quench protocol 2} the tripartite information reaches the asymptotic result \eqref{tripartite domain wall alpha 2}.

\begin{figure}
    \centering
    \includegraphics[width=0.7\textwidth]{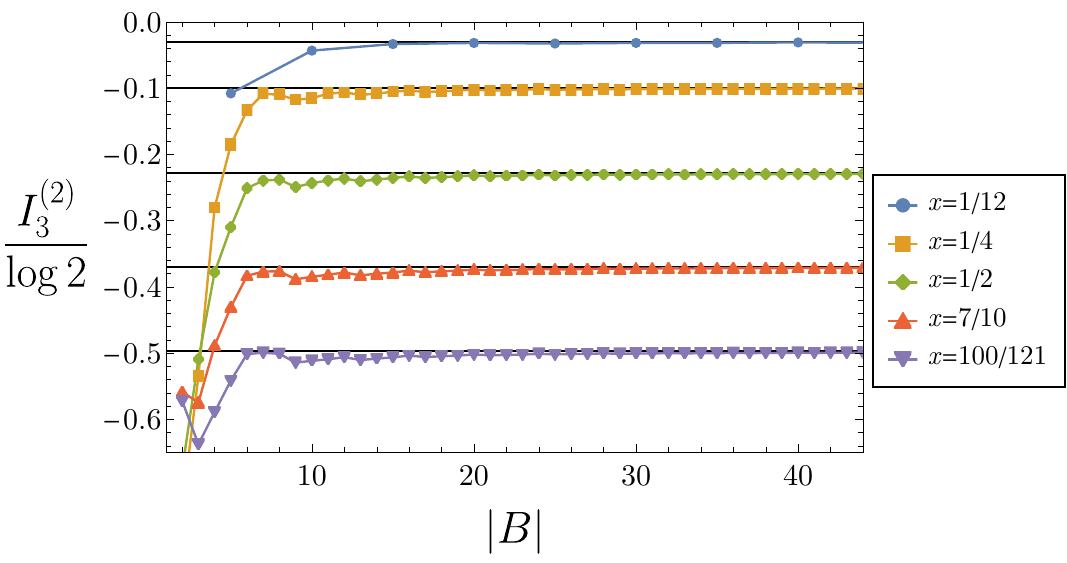}
    \caption{The R\'enyi-$2$ tripartite information in the stationary state of the flipped-spin quench protocol for different configurations with fixed value of the cross ratio $x$, defined by \eqref{eq:cross ratio definition}. From top to bottom the configurations are $(|A|,|B|,|C|)\propto (5,5,1),(1,1,1),(2,1,3),(4,1,7),(10,1,10)$. Increasing the size of the subsystems the tripartite information reaches the analytical prediction \eqref{tripartite domain wall alpha 2}, given by the solid line.}
    \label{fig:xdependence}
\end{figure}

The tripartite information in the stationary state of the flipped-spin protocol exhibits a universal dependency on the cross ratio \eqref{eq:cross ratio definition}, as confirmed by numerical results in figure \ref{fig:xdependence}. The tripartite information is a function only of the cross ratio, notwithstanding that the initial state and the time evolution Hamiltonian are not related to CFT. Note that the tripartite information is zero in the limit $x=0^+$, as in CFT. Differently from CFT, we find nonzero residual tripartite information
\begin{equation}
\textrm{protocol \ref{quench protocol 2}:}\qquad    I_3^{(\alpha)}(A,B,C)\stackrel{x\to 1^-}{=} -\log 2  \; ,
\end{equation}
which corresponds to the limit of small separation between $A$ and $C$ with respect to their size, i.e. to the limit $1\ll |B| \ll |A|,|C|$. The result is the same for all $\alpha$ so it holds also for the von Neumann tripartite information (by replica trick).

\section{Methods}\label{section methods}

\subsection{Kramers-Wannier transformation}\label{section Kramers-Wannier}

The Kramers-Wanniers transformation is fully specified, as discussed in \cite{FagottiMaricZadnik2022}, by the relations
\be\label{eq:dualTD}
\ba
\bs \Pi^{x}_{\tau,-}(\ell)=&\bs\sigma_\ell^x\,,\\
\bs \tau_\ell^y=&\bs \sigma_{\ell-1}^x\bs \sigma_\ell^y\bs\Pi_{\sigma,+}^z(\ell+1)\,,\\
\bs  \tau_\ell^z=&\bs\Pi_{\sigma,+}^z(\ell)\, ,
\ea
\ee
where the operator $\bs\Pi^z_{\sigma,+}(\ell)$ is defined in eq. \eqref{eq:Pi z semilocal operator} and $\bs\Pi^x_{\tau,-}(\ell)$ is defined by
\begin{equation}
    \bs\Pi^{x}_{\tau,-}(\ell)\bs \tau_j^{y,z} =\begin{cases}
       - \bs \tau_j^{y,z} \bs \Pi^{x}_{\tau,-}(\ell) & j\leq \ell \\
          \bs \tau_j^{y,z} \bs \Pi^{x}_{\tau,-}(\ell) & j>\ell   \end{cases}
         , \qquad  \left[\bs \Pi^{x}_{\tau,-}(\ell), \bs\tau_j^x\right]=0, \qquad \left(\bs \Pi^{x}_{\tau,-}(\ell)\right)^2= \bs 1 
\end{equation}
The operators $\bs\Pi^z_{\sigma,+}(\ell)$ and $\bs\Pi^x_{\tau,-}(\ell)$ are semilocal(ized) in the $\sigma$ and $\tau $ basis respectively. They can be thought of as semi-infinite strings $\bs\sigma_\ell^z\bs\sigma_{\ell+1}^z\bs\sigma_{\ell+2}^z\ldots$ and $\ldots \bs\tau_{\ell-2}^x\bs\tau_{\ell-1}^x\bs\tau_{\ell}^x$ respectively. In particular, operators $\bs O$ that are localized in the $\tau$-representation and odd under $\mathcal{P}_\tau^x$ ($\mathcal{P}_\tau^x[\bs O]=-\bs O$) are in the $\sigma$ representation even under $\mathcal{P}_\sigma^z$ ($\mathcal{P}_\sigma^z[\bs O]=\bs O$) and semilocal.

\subsubsection{Semilocal charges}\label{section semilocal charges}
In this section we give a complete set of one-site shift invariant semilocal charges of the dual XY chain \eqref{eq:dual_xy}. A complete set of one-site shift invariant charges of the quantum XY chain \eqref{eq:XY_model} is given by (see e.g. \cite{Fagotti2013Reduced,Fagotti2014On})
\begin{align}
  &  \bs Q^{(0,+)}=\bs H ,\\
  & \bs Q^{(1,+)}=\frac{1}{2} \sum_{\ell=-\infty}^\infty\left[J_x \bs  \tau_\ell^x\bs \tau^x_{\ell+2}+J_y \bs  \tau_\ell^y\bs \tau^y_{\ell+2}-(J_x+J_y)\bs I\right]\bs \tau_{\ell+1}^z,\\
  & \bs Q^{(n,+)}=\frac{1}{2}\sum_{\ell=-\infty}^\infty\left[\left( J_x\bs \tau_{\ell}^x \bs \tau_{\ell+n+1}^x+J_y\bs \tau_{\ell}^y  \bs \tau_{\ell+n+1}^y\right)\prod_{j=1}^n\bs\tau_{\ell+j}^z
  +
  \left( J_x\bs \tau_{\ell}^x \bs \tau_{\ell+n-1}^x+J_y\bs \tau_{\ell}^y  \bs \tau_{\ell+n-1}^y\right)\prod_{j=1}^{n-2}\bs\tau_{\ell+j}^z
  \right],\\
  &\qquad \textrm{for} \quad n=2,3,4,\ldots  \quad ,\nonumber\\
& \label{eq:charge 0 - tau} \bs Q^{(0,-)}=\frac{1}{2}\sum_{\ell=-\infty}^\infty \left(\bs\tau_\ell^x\bs\tau_{\ell+1}^y-\bs\tau_{\ell}^y\bs\tau_{\ell+1}^x\right),\\
&\bs Q^{(n,-)}=\frac{1}{2}\sum_{\ell=-\infty}^\infty \left[\left(\bs\tau_\ell^x\bs\tau_{\ell+n+1}^y-\bs\tau_{\ell}^y\bs\tau_{\ell+n+1}^x\right)\prod_{j=1}^n\bs\tau_{\ell+j}^z\right],
    \qquad n=1,2,3,\ldots
\end{align}
These charges all mutually commute ($[\bs Q^{(n,+)},\bs Q^{(m,\pm)}]=0$ for $n,m\in\mathbb{N}_0$). We note that quantum XY chain possesses also charges that are two-site, and not one-site, shift invariant, which are non-Abelian in general \cite{Fagotti2014On}. These charges are not relevant for the quench protocols considered in this work and will not be discussed. We also note that in interacting systems in general one has to go beyond local charges to consider quasilocal charges \cite{Ilievski2016Quasilocal}, but this is not the case in the studied non-interacting system.

Applying the duality transformation to the charges, we obtain the charges of the dual XY chain \eqref{eq:dual_xy}. While the charges even under $\mathcal{P}_\tau^x$ are local also in the $\sigma$-representation, a simple example being provided by the Hamiltonian, the charges odd under $\mathcal{P}_\tau^x$ are instead semilocal, as discussed in ref. \cite{FagottiMaricZadnik2022} (which labels charges in a different way). Specifically, the charges odd under $\mathcal{P}_\tau^x$ are charges $\bs Q^{(n,+)}$ for odd $n$ and charges $\bs Q^{(n,-)}$ for even $n$. These charges form a complete set of one-site shift invariant semilocal charges of the model in \eqref{eq:dual_xy}. The $\sigma$-representation of the charge $\bs Q^{(0,-)}$ has been already given in eq. \eqref{eq:charge 0 - sigma}. The remaining one-site shift invariant semilocal charges of the model in \eqref{eq:dual_xy} read
\begin{align}
    &\bs  Q^{(1,+)}= \frac{1}{2}\sum_{\ell=-\infty}^\infty\left[ \left(-J_x\bs\sigma_{\ell}^x\bs\sigma_{\ell+1}^y+J_y\bs\sigma_{\ell}^y\bs\sigma_{\ell+1}^x\right)\bs\sigma_{\ell-1}^x\bs\sigma_{\ell+2}^y-(J_x+J_y)\bs\sigma_{\ell+1}^z\bs\sigma_{\ell+2}^z\right]\bs\Pi^z_{\sigma,+}(\ell+3)\\
    &  \bs Q^{(n,+)}=\frac{1}{2}\sum_{\ell=-\infty}^\infty\Bigg[ \left(-J_x \bs\sigma_{\ell}^x \bs\sigma_{\ell+n}^y\prod_{j=1}^{(n-1)/2}\bs \sigma_{\ell+2j-1}^z +J_y \bs\sigma_{\ell}^y \bs\sigma_{\ell+n}^x\prod_{j=1}^{(n-1)/2}\bs \sigma_{\ell+2j}^z\right)\bs \sigma_{\ell-1}^x \bs\sigma_{\ell+n+1}^y + \nonumber \\ 
& \left(-J_x \bs\sigma_{\ell}^x \bs\sigma_{\ell+n-2}^y\prod_{j=1}^{(n-3)/2}\bs \sigma_{\ell+2j-1}^z +J_y \bs\sigma_{\ell}^y \bs\sigma_{\ell+n-2}^x\prod_{j=1}^{(n-3)/2}\bs \sigma_{\ell+2j}^z\right)\bs \sigma_{\ell-1}^x \bs\sigma_{\ell+n-1}^y  \bs\sigma_{\ell+n}^z  \bs\sigma_{\ell+n+1}^z 
       \Bigg]\bs\Pi^z_{\sigma,+}(\ell+n+2) \label{eq:charge n + sigma}\\
& \qquad \mathrm{for} \quad n=3,5,7,\ldots  \quad ,\nonumber\\
&\bs Q^{(n,-)}=\frac{1}{2}\sum_{\ell=-\infty}^\infty\left(\bs\sigma_{\ell}^x\bs\sigma_{\ell+n}^x\prod_{j=1}^{n/2}\bs\sigma_{\ell+2j-1}^z+\bs\sigma_{\ell}^y\bs\sigma_{\ell+n}^y\prod_{j=1}^{n/2-1}\bs\sigma_{\ell+2j}^z\right)\bs\sigma_{\ell-1}^x\bs\sigma_{\ell+n+1}^y\bs\Pi_{\sigma,+}^z(\ell+n+2) \\ 
&\qquad \mathrm{for} \quad n=2,4,6,\ldots  \quad, \nonumber 
\end{align}
where we use the convention that products of the form $\prod_{j=1}^0$ are equal to identity.

\subsubsection{Semilocal Generalized Gibbs ensemble}

As pointed out in \cite{FagottiMaricZadnik2022}, the stationary state of quench protocol \ref{quench protocol 1} is a \textit{semilocal} generalized Gibbs ensemble, meaning that it includes semilocal charges. Here we point out that the non-equilibrium stationary state emerging from the flipped-spin quench protocol is also described by a semilocal GGE.

In both cases the GGE can be obtained directly by working in the dual picture and following the approach of ref. \cite{Fagotti2013Reduced}, as explained in appendix \ref{appendix Lagrange multipliers}.
We find that the GGE
\begin{equation}\label{GGE XY general}
    \bs\rho=\frac{1}{\mathcal{Z}}\exp\left[-\sum_{n=0}^\infty \left(\lambda^{(n,+)}\bs Q^{(n,+)}+\lambda^{(n,-)}\bs Q^{(n,-)}\right) \right] \; ,
\end{equation}
where $\mathcal{Z}$ is the normalization, is specified by Lagrange multipliers
\begin{align} \label{lagrange multiplier protocol 1}
       \textrm{protocol \ref{quench protocol 1}:}\qquad & \lambda^{(n,-)}=0, \qquad \lambda^{(n,+)}=\begin{cases}
           0, & \textrm{$n$ even}\\
          \frac{8}{\pi} \int_0^{\pi/2} \mathrm{artanh}\left[\frac{2(J_x+J_y)\cos k}{\varepsilon(k)}\right]\frac{1}{\varepsilon(k)}\cos(nk)\ dk ,&  \textrm{$n$ odd}
       \end{cases}, \\
\textrm{protocol \ref{quench protocol 2}:} \qquad & \lambda^{(n,-)}=\begin{cases}
    \mathrm{sgn}(J_xJ_y)\frac{8}{\pi} \int_0^{\pi/2} \mathrm{artanh}\left[\frac{2(J_x+J_y)\cos k}{\varepsilon(k)}\right]\sin[(n+1)k]\ dk , & \textrm{$n$ even}\\
    0, & \textrm{n odd}
\end{cases}  ,\quad \lambda^{(n,+)}=0, \quad  \label{lagrange multiplier protocol 2}
\end{align}
for $n\in\mathbb{N}_0$.

The Lagrange multipliers are completely different in the two quench protocols, but in both cases only those associated to semilocal charges are nonzero. In the dual picture these are the charges odd under $\mathcal{P}_\tau^x$, which will be important, in section \ref{section invariance}, for arguing that the tripartite information is not influenced by the Kramers-Wannier transformation. We note that the nonzero Lagrange multipliers decay rather slowly with $n$, i.e. only algebraically, which is related to the logarithmic singularity of the function $\mathrm{artanh}[2(J_x+J_y)/\varepsilon(k)]$, appearing under the integral, at $k=0$.

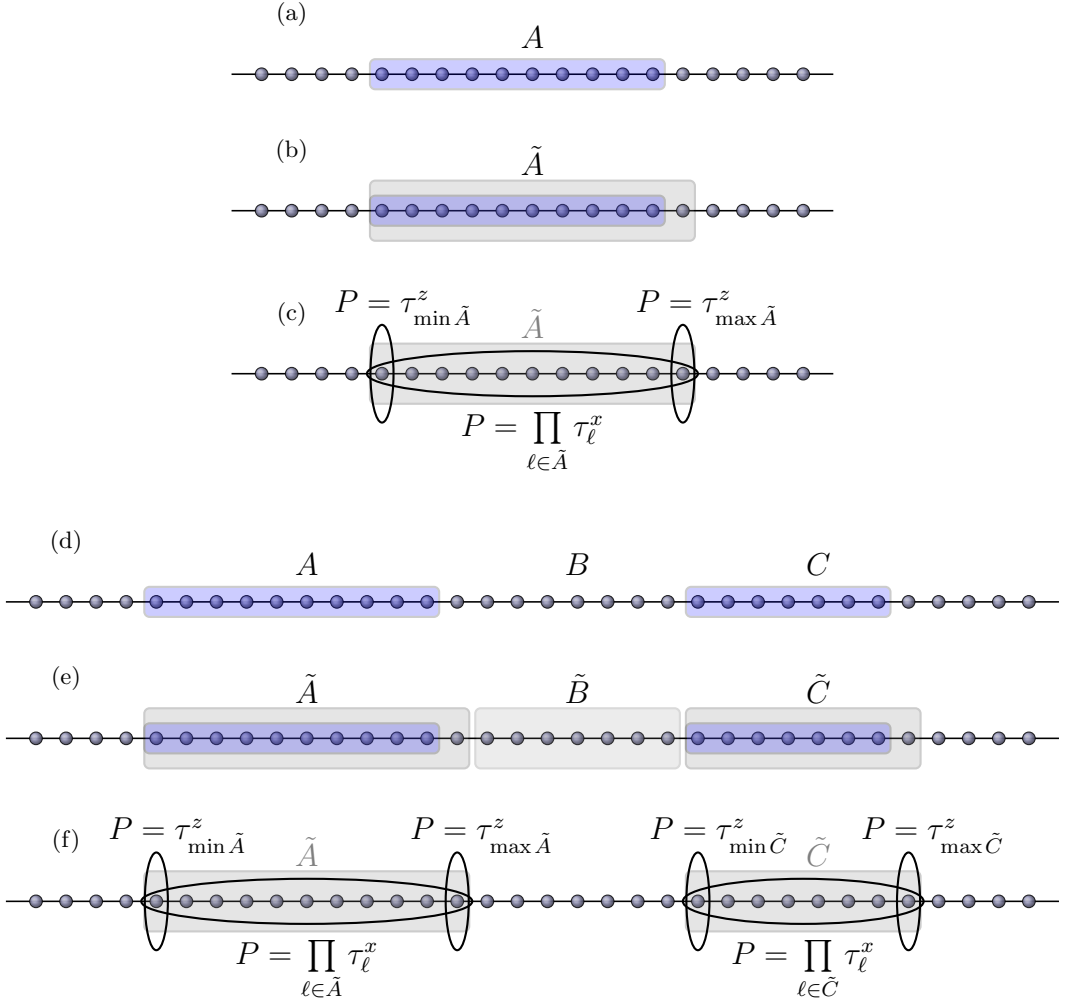
\begin{figure}

\begin{tikzpicture}[scale=0.4,every node/.style={scale=1}]
    	\node[anchor=center] at (1,2) {(a)};
    \draw[black,line width=0.6pt] (-1,0) to (19,0);
    \foreach \x in {0,...,18}
    \filldraw[ball color=blue!20!white,opacity=0.75,shading=ball] (\x,0) circle (6pt);
    \node[anchor=south] at (9,0.6) {\large $ A$};
 \draw[fill=blue,opacity=0.2, rounded corners = 2,thick] (3.6,-0.5) rectangle ++(9.8,1);
    \end{tikzpicture}
    \\
\vspace{0.5 cm}
 \begin{tikzpicture}[scale=0.4]
    	\node[anchor=center] at (1,2) {(b)};
    \draw[black,line width=0.6pt] (-1,0) to (19,0);
    \foreach \x in {0,...,18}
    \filldraw[ball color=blue!20!white,opacity=0.75,shading=ball] (\x,0) circle (6pt);
    \node[anchor=south] at (9,0.9) {\large $\tilde A$};
    \draw[rounded corners = 2,thick,fill=gray,opacity=0.2] (3.6,-1) rectangle ++(10.8,2);
 \draw[fill=blue,opacity=0.2, rounded corners = 2,thick] (3.6,-0.5) rectangle ++(9.8,1);
    \end{tikzpicture}
\\
\vspace{0.5 cm}
    \begin{tikzpicture}[scale=0.4]
    	\node[anchor=center] at (1,2) {(c)};
    \draw[black,line width=0.6pt] (-1,0) to (19,0);
    \foreach \x in {0,...,18}
    \filldraw[ball color=blue!20!white,opacity=0.75,shading=ball] (\x,0) circle (6pt);
    \node[anchor=south,text=gray] at (9,0.9) {\large $\tilde A$};
    \draw[rounded corners = 2,thick,fill=gray,opacity=0.2] (3.6,-1) rectangle ++(10.8,2);
\node[ellipse,draw,thick,minimum width = 4.4 cm, 
    minimum height = 0.6 cm] (e) at (9,0) {};
    \node[ellipse,draw,thick,minimum width = 0.25 cm, 
    minimum height = 1.3 cm] (e) at (4,0) {};
     \node[ellipse,draw,thick,minimum width = 0.25 cm, 
    minimum height = 1.3 cm] (e) at (14,0) {};
    \node[anchor=center] at (4.7,2.2) {\large $P=\tau_{\mathrm{min}\ \!\! \tilde A}^z$};
    \node[anchor=center] at (14.8,2.2) {\large $P=\tau_{\mathrm{max}\ \!\! \tilde A}^z$};
    \node[anchor=center] at (9,-2.3) {\large $P=\prod\limits_{\ell\in \tilde A}\tau_\ell^x$};
    \end{tikzpicture}
   \\
\vspace{0.5 cm}
 \begin{tikzpicture}[scale=0.4]
    	\node[anchor=center] at (1,2) {(d)};
    \draw[black,line width=0.6pt] (-1,0) to (34,0);
    \foreach \x in {0,...,33}
    \filldraw[ball color=blue!20!white,opacity=0.75,shading=ball] (\x,0) circle (6pt);
    \node[anchor=south] at (9,0.6) {\large $ A$};
     \node[anchor=south] at (26.,0.6) {\large $ C$};
\node[anchor=south] at (18,0.6) {\large $ B$};
       \draw[fill=blue,opacity=0.2, rounded corners = 2,thick] (3.6,-0.5) rectangle ++(9.8,1);
        \draw[fill=blue,opacity=0.2, rounded corners = 2,thick] (21.6,-0.5) rectangle ++(6.8,1);
    
    \end{tikzpicture}
    \\
\vspace{0.5 cm}
 \begin{tikzpicture}[scale=0.4]
    	\node[anchor=center] at (1,2) {(e)};
    \draw[black,line width=0.6pt] (-1,0) to (34,0);
    \foreach \x in {0,...,33}
    \filldraw[ball color=blue!20!white,opacity=0.75,shading=ball] (\x,0) circle (6pt);
    \node[anchor=south] at (9,0.9) {\large $\tilde A$};
     \node[anchor=south] at (26.,0.9) {\large $\tilde C$};
     \node[anchor=south] at (18,0.9) {\large $\tilde B$};
    \draw[rounded corners = 2,thick,fill=gray,opacity=0.2] (3.6,-1) rectangle ++(10.8,2);
      \draw[rounded corners = 2,thick,fill=gray,opacity=0.2] (21.6,-1) rectangle ++(7.8,2);
\draw[rounded corners = 2,thick,fill=gray,opacity=0.15] (14.6,-1) rectangle ++(6.8,2);

       \draw[fill=blue,opacity=0.2, rounded corners = 2,thick] (3.6,-0.5) rectangle ++(9.8,1);
        \draw[fill=blue,opacity=0.2, rounded corners = 2,thick] (21.6,-0.5) rectangle ++(6.8,1);
    
    \end{tikzpicture}
\\
\vspace{0.5 cm}
 \begin{tikzpicture}[scale=0.4]
    	\node[anchor=center] at (1,2) {(f)};
    \draw[black,line width=0.6pt] (-1,0) to (34,0);
    \foreach \x in {0,...,33}
    \filldraw[ball color=blue!20!white,opacity=0.75,shading=ball] (\x,0) circle (6pt);
    \node[anchor=south,text=gray] at (9,0.9) {\large $\tilde A$};
     \node[anchor=south,text=gray] at (26.,0.9) {\large $\tilde C$};
    \draw[rounded corners = 2,thick,fill=gray,opacity=0.2] (3.6,-1) rectangle ++(10.8,2);

      \draw[rounded corners = 2,thick,fill=gray,opacity=0.2] (21.6,-1) rectangle ++(7.8,2);

\node[ellipse,draw,thick,minimum width = 4.4 cm, 
    minimum height = 0.6 cm] (e) at (9,0) {};
    \node[ellipse,draw,thick,minimum width = 0.25 cm, 
    minimum height = 1.3 cm] (e) at (4,0) {};
     \node[ellipse,draw,thick,minimum width = 0.25 cm, 
    minimum height = 1.3 cm] (e) at (14,0) {};

\node[ellipse,draw,thick,minimum width = 3.2 cm, 
    minimum height = 0.6 cm] (e) at (25.5,0) {};
    \node[ellipse,draw,thick,minimum width = 0.25 cm, 
    minimum height = 1.3 cm] (e) at (22,0) {};
     \node[ellipse,draw,thick,minimum width = 0.25 cm, 
    minimum height = 1.3 cm] (e) at (29,0) {};
    
    \node[anchor=center] at (4.7,2.2) {\large $P=\tau_{\mathrm{min}\ \!\! \tilde A}^z$};
    \node[anchor=center] at (14.8,2.2) {\large $P=\tau_{\mathrm{max}\ \!\! \tilde A}^z$};
    \node[anchor=center] at (9,-2.3) {\large $P=\prod\limits_{\ell\in \tilde A}\tau_\ell^x$};

\node[anchor=center] at (22.7,2.2) {\large $P=\tau_{\mathrm{min}\ \!\! \tilde C}^z$};
    \node[anchor=center] at (29.8,2.2) {\large $P=\tau_{\mathrm{max}\ \!\! \tilde C}^z$};
    \node[anchor=center] at (25.5,-2.3) {\large $P=\prod\limits_{\ell\in \tilde C}\tau_\ell^x$};
    
    \end{tikzpicture}

 \caption{(a) A connected block $A$. (b) Block $\tilde A$ we work with in the dual picture, related to the configuration in a). (c) The transformations $P_j$ appearing in eq.~\eqref{rdm connected block}. (d) Disjoint blocks $A$ and $C$. (e) Blocks $\tilde A, \tilde B, \tilde C$ we work with in the dual picture, related to the configuration in d). (f) The transformations $P_j$ appearing in eq.~\eqref{rdm as an average}.}
    \label{fig: blocks and transformations}
    
\end{figure}

\subsubsection{The reduced density matrix in the dual picture}\label{section reduced density matrix in the dual picture}

In this section we find a convenient representation of the reduced density matrix \eqref{density matrix sigma basis tensor} in view of comparing it with \eqref{density matrix tau basis tensor}. For this task we introduce the Kramers-Wannier transformation for a finite system. Let us consider a system of size $L$, described by Pauli matrices $\sigma^\gamma_\ell\equiv \mathbb I^{\otimes (\ell-1)}\otimes\sigma^\gamma \otimes \mathbb I^{\otimes (L-\ell)}$ for $\gamma=0,x,y,z$ and $\ell=1,2,\ldots,L$, where $\sigma^0=\mathbb{I}$ is a $2\times 2$ unit matrix. The transformation is given by
\be\label{duality finite system even}
\tau_\ell^x=\sigma_{\ell-1}^x\sigma_{\ell}^x\, ,\qquad
\tau_{\ell'}^z\tau_{\ell'+1}^z=\sigma_{\ell'}^z\, ,
\ee
for $\ell=2,3,\ldots,L$ and $\ell'=1,2,\ldots,L-1$. It is in correspondence with the one in \eqref{eq:dualTDeven} for the infinite system. Eq. \eqref{duality finite system even} does not specify the transformation of operators $O$ that are odd under $\Pi^z_\sigma\equiv\prod_{\ell=1}^L\sigma_\ell^z$ ($[O,\Pi^z_\sigma]=-O$) nor the transformation at the boundaries. The full mapping can be found in \cite{Fagotti2022Global}. For our purposes it is just important that the mapping preserves the algebra of Pauli matrices and that for operators even under $\Pi^z_\sigma$ that are not at the boundaries the mapping is analogous to the one for the infinite system. Note that to describe a connected block $A$ we need to consider the mapping \eqref{duality finite system even} for $L\geq |A|+1$, to avoid the peculiarities of the mapping at the boundaries .

For a system in state $\ket{\Psi}$, the reduced density matrix for subsystem $X$ is given by \eqref{density matrix sigma basis tensor}. In order to define a duality transformation in correspondence with the one in the infinite system we consider an enlarged space $\tilde X=X\cup X'$, for some suitable choice of $X'$ that will be discussed later. Note that we can extend $\rho_X$ by identity to the enlarged space, defining
\begin{equation}\label{enlarged density matrix}
    \bar \rho_{X}\equiv \rho_X\otimes\left( \bigotimes\limits_{\ell\in X'}\frac{\mathbb I}{2}\right) \; .
\end{equation}
To obtain the entanglement entropy from the enlarged density matrix \eqref{enlarged density matrix} we just need to remove a constant related to the size of the enlargement $X'$,
\begin{equation}\label{entropy enlargement of the space}
    S_\alpha(X)=\frac{1}{1-\alpha}\log\tr (\bar \rho_{X}^\alpha)-|X'|\log 2 \; .
\end{equation}
Having the duality transformation defined in the enlarged space, it is convenient to consider the density matrices
\begin{equation}\label{eq:rdm rho A}
   \tilde \rho_{ Y}=\frac{1}{2^{|\tilde X|}}\sum_{\gamma_{\ell}\in\{0,x,y,z\},\ell\in Y} \bra{\Psi}\prod_{\ell\in Y}\bs \tau^{\gamma_\ell}_\ell \ket{\Psi}\prod_{\ell\in Y}\tau_\ell^{\gamma_\ell} \; .
\end{equation}
for subsystems $Y\subseteq \tilde X$. Clearly, the density matrix \eqref{eq:rdm rho A} is in correspondence with the density matrix $\rho^\tau_{Y}$, defined by \eqref{density matrix tau basis tensor}, but they are different mathematical objects. Instead of comparing the density matrices \eqref{density matrix sigma basis tensor} and \eqref{density matrix tau basis tensor} directly we will find it more convenient to compare the density matrices \eqref{enlarged density matrix} and \eqref{eq:rdm rho A}, for a suitable choice of $Y$, concluding about the former afterwards.

\paragraph{Single block}
When the subsystem  $X$ is a connected block $A$ (figure \ref{fig: blocks and transformations}a) we consider the enlarged subsystem $\tilde A=A\cup \{\mathrm{max}(A)+1\}$ (figure \ref{fig: blocks and transformations}b). There is a simple representation for $\bar\rho_A=\rho_A\otimes \frac{\mathbb I}{2}$ in the dual picture,
\begin{equation}\label{rdm connected block}
    \bar\rho_A = \frac{1}{2^3}\sum_{j_1,j_2,j_3=0}^1 \left(\prod_{m=1}^3{P_m^{j_m}} \right)\tilde \rho_{\tilde A}\left(\prod_{m=1}^3{P_m^{j_m}}\right)^\dagger ,
\end{equation}
with
\begin{equation} 
     P_1=\prod_{\ell\in \tilde A}\tau_\ell^x  ,\qquad       P_2=\tau_{\mathrm{min}(\tilde A)}^z , \qquad
       P_3 =\tau_{\mathrm{max}(\tilde A)}^z .
\end{equation}
This result has already been obtained in \cite{FagottiMaricZadnik2022} and we revisit it here for comparison with the case of disjoint blocks. The density matrix $\bar\rho_{A}$ is obtained from $\tilde \rho_{\tilde A}$ by successive transformations of the form $\rho\to (\rho +P\rho P)/2$, for three different Hermitian involutions $P$. Their effect is to remove those terms from $\tilde \rho_{\tilde A}$ which act non-trivially on the site $\mathrm{max}(A)+1$ (in the $\sigma$-representation), leaving only terms with support on $A$. Note that transformation $P_1$ has support on the whole system $\tilde A$ and acts by flipping all $\tau_\ell^z$, while the remaining transformations affect only the boundaries. All three transformations are represented graphically in figure \ref{fig: blocks and transformations}c.

The representation \eqref{rdm connected block} follows from the following properties, that are direct consequences of the Kramers-Wannier transformation \eqref{eq:dualTDeven}:
\begin{enumerate}[1)]
    \item\label{property 1} Any operator $\bs O_A$ localized (in the $\sigma$ representation) in $A$ and even under $\mathcal{P}_\sigma^z$ is mapped by the Kramers-Wannier transformation to an operator that in the $\tau$ representation is localized in $\tilde A$ and is even under $\mathcal{P}_\tau^x$ ($\mathcal{P}_\tau^x[\bs O_A]=\bs O_A$). More technically, if $\bs O_A=\prod_{\ell\in A}\bs\sigma_\ell^{\gamma_\ell}$, where $\gamma_\ell\in\{0,x,y,z\}$, is even under $\mathcal{P}_\sigma^z$ then we have $\bs O_A=s\prod_{\ell'\in \tilde A}\bs\tau_{\ell'}^{\gamma'_{\ell'}}$, for some $\gamma_{\ell'}'\in\{0,x,y,z\}$, with $s\in\{\pm1\} $ allowing for a minus sign.

    \item \label{property 2}Any operator $\bs O_{\tilde A}=\prod_{\ell\in \tilde A}\bs\tau_{\ell}^{\gamma_\ell}$, where $\gamma_\ell\in\{0,x,y,z\}$, that is even under $\mathcal{P}_\tau^x$ and which in the $\sigma$ representation has support outside $A$, anticommutes with $\bs \sigma_{\min(A)-1}^z$ ($\{\bs O_{\tilde A}, \bs \sigma_{\min(A)-1}^z\}=0$) or $\bs \sigma_{\max(A)+1}^z$ ($\{\bs O_{\tilde A}, \bs \sigma_{\max(A)+1}^z\}=0$), or both. 
\end{enumerate}
A simple example of property \ref{property 2} are operators $\bs \tau_{\min(\tilde A)}^x=\bs \sigma_{\min(A)-1}^x \bs \sigma_{\min(A)}^x$ and $\bs \tau^x_{\max(\tilde A)}=\bs \sigma^x_{\max(A)}\bs \sigma^x_{\max(A)+1}$.

\paragraph{Disjoint blocks}
When the subsystem $X$ consists of disjoint blocks $A$ and $C$ (figure \ref{fig: blocks and transformations}d) we don't have such a simple relation as \eqref{rdm connected block}. The additional complication is that the duality transformation can introduce strings between the blocks. For example,
\begin{equation}\label{string x example}
    \bs\sigma_{\max(A)}^x \bs\sigma_{\min(C)}^x=\prod_{\ell=\max (A)+1}^{\min (C)} \bs \tau_\ell^x .
\end{equation}
These strings are similar to the ones appearing in expressing the reduced density matrix of disjoint blocks in terms of Jordan-Wigner fermions \cite{Fagotti2010disjoint} and we tackle the problem in a similar way.

We have to consider the enlargement of $X=A\cup C$ to the whole subsystem between $A$ and $C$, and to the site adjacent to $C$ on the right. We thus work with $X'=\{\max(A)+1,\max(A)+2,\ldots, \min(C)-1\}\cup \{\max(C)+1\}$. It is also convenient to define the subsystems $\tilde A\equiv A\cup \{\max(A)+1\}$, $\tilde B\equiv B-\{\max(B)\}$ and $\tilde C\equiv C\cup \{\max(C)+1\}$ (figure \ref{fig disjoint blocks}e). Here we assume $|B|\geq 2$. Similarly to the single block case, we consider the density matrix \eqref{eq:rdm rho A} with $Y=\tilde A\tilde C$, which can be written as
\begin{equation}\label{rdm part without string rho}
    \tilde \rho_{\tilde A\tilde C} = \frac{1}{2^{|\tilde A\tilde B \tilde C|}}\sum_{O_{\tilde A}, O_{\tilde C}}\bra{\Psi}\bs{ O}_{\tilde A} \bs{ O}_{\tilde C} \ket{\Psi} O_{\tilde A}O_{\tilde C} \; ,
\end{equation}
where the sum is over all possible products $O_{\tilde A}=\prod_{\ell\in \tilde A}\tau_\ell^{\gamma_\ell}$ with $\gamma_\ell\in\{0,x,y,z\}$ and $O_{\tilde C}$, defined analogously for subsystem $\tilde C$.
The problem of density matrix \eqref{rdm part without string rho} is that it cannot describe operators such as \eqref{string x example}, that include strings between the blocks in the dual picture. To describe them we introduce the traceless operator
   \begin{equation}\label{rdm string omega}
        \omega_{\tilde A \tilde C}=\frac{1}{2^{|\tilde A\tilde B \tilde C|}}\sum_{O_{\tilde A}, O_{\tilde C}} \bra{\Psi}\bs{ O}_{\tilde A} \bs S^x \bs{O}_{\tilde C} \ket{\Psi} (O_{\tilde A}O_{\tilde C})S^x,
    \end{equation}
which is similar to $\tilde \rho_{\tilde A\tilde C}$, but includes the string $S^x\equiv\prod_{\ell\in\tilde B}\tau_\ell^x$ ($\bs S^x\equiv\prod_{\ell\in\tilde B}\bs \tau_\ell^x$).

We find that the reduced density matrix $\bar\rho_{AC}=\rho_{AC}\otimes( \bigotimes_{\ell\in B \cup \{\max(C)+1\}}\mathbb I/2)$ is equal to
\begin{equation}\label{rdm as an average}
\begin{split}
    \bar\rho_{AC} = & \frac{1}{2^6}\sum_{j_1,j_2, \ldots ,j_6=0}^1 \left( \prod_{m=1}^6{P_m^{j_m}} \right) \tilde \rho_{\tilde A\tilde C}\left(\prod_{m=1}^6{P_m^{j_m}}\right)^\dagger\\+ &\frac{1}{2^6}\sum_{j_1,j_2, \ldots ,j_6=0}^1 (-1)^{j_4+j_5}\left( \prod_{m=1}^6{P_m^{j_m}} \right)\omega_{\tilde A\tilde C} \left(\prod_{m=1}^6{P_m^{j_m}}\right)^\dagger 
\end{split}
\end{equation}
where
\begin{equation}
\begin{split}
     &P_1=\prod_{\ell\in \tilde A}\tau_\ell^x  ,\qquad       P_2=\tau_{\mathrm{min}(\tilde A)}^z , \qquad
       P_3 =\tau_{\mathrm{max}(\tilde A)}^z ,\\
      & P_4=\prod_{\ell\in \tilde C}\tau_\ell^x  ,\qquad       P_5=\tau_{\mathrm{min}(\tilde C)}^z , \qquad
       P_6 =\tau_{\mathrm{max}(\tilde C)}^z .
\end{split}
\end{equation}
The transformations are completely analogous to the single block case and are graphically represented in figure \ref{fig: blocks and transformations}f.

The derivation of representation \eqref{rdm as an average} is more pedantic than in the single block case, but similar, so we cover just the main ingredients. It is derived by applying properties \ref{property 1} and \ref{property 2} to blocks $A$ and $C$ separately. The additional complication is that the non-locality of the transformation can introduce strings between the blocks, as in \eqref{string x example}. To tackle it we notice that any product of Pauli matrices $\bs O_X$ localized (in the $\sigma$ representation) in $X=A\cup C$ and even under $\mathcal{P}_\sigma^z$ ($\mathcal{P}_\sigma^z[\bs O_X]=\bs O_X$) can be written as a product $\bs O_X=\bs O_A\bs O_C$ where $\bs O_A$ and $\bs O_C$ are localized in $A$ and $C$ respectively, and have the same parity under $\mathcal{P}_\sigma^z$, i.e. we have that either $\bs O_A$ and $\bs O_C$ are both even under $\mathcal{P}_\sigma^z$ ($\mathcal{P}_\sigma^z[\bs O_{A,C}]=\bs O_{A,C}$) or they are both odd ($\mathcal{P}_\sigma^z[\bs O_{A,C}]=-\bs O_{A,C}$).
The even operators $\bs O_A$ and $\bs O_C$ are, by property \ref{property 1}, localized in $\tilde A$ and $\tilde C$ respectively so an analogous application of property \ref{property 2} to the single block case results in the first term in \eqref{rdm as an average}. The second term is a contribution from the terms where both $\bs O_A$ and $\bs O_C$ are odd. To treat such terms we notice that they can be rewritten as $\bs O_A=(\bs O_A \bs \sigma_{\ell_A}^x) \bs \sigma_{\ell_A}^x$ and $\bs O_C=(\bs O_C \bs \sigma_{\ell_C}^x) \bs \sigma_{\ell_C}^x$, where $\ell_A,\ell_C$ is an arbitrary site belonging to $A,C$ respectively. The operators in the brackets are even under $\mathcal{P}_\sigma^z$ so they can be dealt with by applying properties \ref{property 1} and \ref{property 2}, while the product $\bs \sigma_{\ell_A}^x\bs \sigma_{\ell_C}^x=\bs\tau_{\ell_A+1}^x\bs \tau_{\ell_A+2}^x\ldots \bs \tau_{\ell_C}^x$ is a string, common to all odd $\bs O_A$, $\bs O_C$. This gives the second term in \eqref{rdm as an average}. A subtlety is that the sign factor $(-1)^{j_4+j_5}$ appears, because the operators $\bs \sigma^z_{\max (A)}=\bs\tau^z_{\max (\tilde A)}\bs\tau^z_{\max (\tilde A)+1}$ and $\bs \sigma^z_{\min(\tilde C)-1}=\bs\tau^z_{\min (\tilde C)-1}\bs\tau^z_{\min(\tilde C)}$ arising from property \ref{property 2} anticommute with the string $\bs S^x$.

\subsubsection{Invariance of the tripartite information}\label{section invariance}

Based on representations \eqref{rdm connected block} and \eqref{rdm as an average} for the reduced density matrices we now argue that the tripartite information in the stationary states of the studied protocols is not influenced by the Kramers-Wannier transformation.

The first thing to notice is that, in the model \eqref{eq:XY_model}, that does not have semilocal charges, the string order does not survive the quench (see \cite{FagottiMaricZadnik2022} for a general discussion){, i.e. in the $\tau$ representation there is no string order after the quench}. Thus, since $\omega_{\tilde A \tilde C}$, defined in \eqref{rdm string omega}, includes strings between the blocks {(in the $\tau$ representation)}, in the limit of large subsystems it is expected to be negligible with respect to $\tilde \rho_{\tilde A\tilde C}$. Therefore, the second line in the expression \eqref{rdm as an average} for the reduced density matrix of disjoint blocks in the dual picture can be neglected, leaving us with 
\begin{equation}
    \bar\rho_{AC} \sim \frac{1}{2^6}\sum_{j_1,j_2, \ldots ,j_6=0}^1 \left( \prod_{m=1}^6{P_m^{j_m}} \right) \tilde \rho_{\tilde A\tilde C}\left(\prod_{m=1}^6{P_m^{j_m}}\right)^\dagger,
\end{equation}
where $\sim$ stands for the limit $|A|,|B|,|C|\gg 1$.

The transformations $P$ acting on the boundaries act in the same way in the single block case and in the case of disjoint blocks (compare Fig. \ref{fig: blocks and transformations}c and Fig. \ref{fig: blocks and transformations}f). Moreover, when computing different entropies appearing in the definition \eqref{eq:tripartite definition in terms of S} of the tripartite information, these boundary transformations always act (in the dual picture) on the sites $\ell =\min A,\max{\tilde A}, \min{\tilde C},\max{\tilde C}$. For example, when computing $S_\alpha(B)$ we have transformations that act on the sites $\ell=\min B=\max \tilde A $ and $\ell=\max(B)+1=\min{\tilde C}$. Since the tripartite information is constructed in such a way to cancel the contributions of the boundaries, these transformations are not expected to affect it. Therefore, the same tripartite information should be obtained by starting from the density matrix
\begin{equation}\label{step single block rho prime}
    \bar\rho_A' \equiv \frac{1}{2}\sum_{j=0}^1 P_3^{j}\tilde \rho_{\tilde A} P_3^{j} 
\end{equation}
for the reduced density matrix of a single block and
\begin{equation}\label{step disjoint rho prime}
    \bar\rho_{AC}' \equiv \frac{1}{2^2}\sum_{j_3,j_6=0}^1 \left( \prod_{m=3,6}{P_m^{j_m}} \right) \tilde \rho_{\tilde A\tilde C}\left(\prod_{m=3,6}{P_m^{j_m}}\right)
\end{equation}
for the reduced density matrix of disjoint blocks.

The stationary state of both of the studied quench protocols can be written as
\begin{equation}
    \bs \rho_{\mathrm{GGE}}=\frac{1}{\mathcal{Z}}e^{-\bs Q},
\end{equation}
where $\mathcal{Z}$ is the normalization and $\bs Q$ includes the relevant charges with appropriate Lagrange multipliers, given by eq. \eqref{lagrange multiplier protocol 1} and eq. \eqref{lagrange multiplier protocol 2}. The operator $\bs Q$ can be interpreted as a Hamiltonian and its eigenvalues $E_i$ as energies, that come in pairs $\pm E_i$. Since the GGE entropy is extensive with the system size, the energies $E_i$ are also expected to be extensive. It follows that for any integer $M$ such that $1\leq M\leq \alpha-1$ the ratios
\begin{equation}\label{step cosh charge energy}
   \frac{\tr\left[e^{ -(\alpha-2M)\bs Q}\right]}{\tr\left[e^{-\alpha\bs Q}\right]} =  \frac{\sum_{E_i\geq 0}\cosh[(\alpha-2M)E_i]}{\sum_{E_i\geq 0} \cosh(\alpha E_i)}
\end{equation}
are exponentially suppressed with the system size, while for $M=0, \alpha$ they are equal to unity.

The reduced density matrices $\tilde\rho_{\tilde A}$ and $\tilde \rho_{\tilde A\tilde C}$ are obtained by tracing out the GGE density matrix in the dual picture. Let us focus on the single block case. When the block $\tilde A$ is large enough the reduced density matrix $\tilde\rho_{\tilde A}$ is expected to be similar to the GGE in the bulk. For thermal states the entanglement Hamiltonian is at large temperatures approximated well by the Hamiltonian of the subsystem \cite{Dalmonte2022}. Similarly, we expect that the properties of the entanglement Hamiltonian in the GGE can be captured well by the restriction of $\bs Q$ to the subsystem.
The consequence is that for $j_1,\ldots, j_{\alpha-1}\in\{0,1\}$, such that $M$ of them are non-zero, with $1<M<\alpha$, we have
\begin{equation}\label{step trace rho P }
    \frac{\tr\left[ \rho \left( P_{m_1}^{j_1}\rho P^{j_1}_{m_1}\right) \left(P^{j_2}_{m_2}\rho P^{j_2}_{m_2} \right)\ldots \left(P^{j_{\alpha-1}}_{m_{\alpha-1}}\rho P^{j_{\alpha-1}}_{m_{\alpha-1}}\right)\right]}{\tr \rho^\alpha} \sim 0 \; ,
\end{equation}
where $m_1,\ldots,m_{\alpha-1}=3$. Here we have used the property that $\bs Q$ includes only charges odd under $\mathcal{P}^\tau_x$, realization that the transformation $\bullet \to P_3\bullet P_3$ is a finite-size analogue of transformation $\mathcal{P}^\tau_x$, which changes the sign in front of the odd charges, and the suppression of the ratios \eqref{step cosh charge energy}. In the case of disjoint blocks, the transformations $P_3,P_6$, appearing in \eqref{step disjoint rho prime}, act only on a part of the subsystem, but their support is still extensive with the size of one of the subsystems and we expect the ratios in \eqref{step trace rho P } with $m_1,\ldots,m_{\alpha-1}\in\{3,6\}$ to be suppressed for similar reasons.

From \eqref{step trace rho P } it follows
\begin{equation}\label{step removing P}
    \tr\left[ \left(\bar \rho_{A}'\right)^\alpha\right]\sim 2^{1-\alpha} \tr\left[ \left(\tilde \rho_{\tilde A}\right)^\alpha\right], \qquad  \tr\left[ \left(\bar \rho_{AC}'\right)^\alpha\right]\sim 2^{2(1-\alpha)} \tr\left[ \left(\tilde \rho_{\tilde A\tilde C}\right)^\alpha\right] \; ,
\end{equation}
where in the disjoint blocks case the squared prefactor appears because in eq. \eqref{step disjoint rho prime} we have two transformations. Now, since the prefactors in \eqref{step removing P} cancel in the tripartite information it follows that the same tripartite information is obtained if we use $\tilde \rho_{\tilde A}$ instead of $\bar\rho_A$ (analogously for other single block contributions to the tripartite information in \eqref{eq:tripartite definition in terms of S}) and $\tilde \rho_{\tilde A\tilde C}$ instead of $\bar\rho_{AC}$. Since $\tilde \rho_{\tilde A},\tilde \rho_{\tilde A\tilde C}$ are in correspondence with the density matrices $\rho_{\tilde A}^\tau,\rho_{\tilde A\tilde C}^\tau$, defined by \eqref{density matrix tau basis tensor}, we can simply use the latter as well. The conclusion is that for the stationary states of the studied quench protocols the tripartite information is not influenced by the Krammers-Wannier transformation: the differences in the R\'enyi entropies arising from the differences between the density matrices \eqref{density matrix sigma basis tensor} and \eqref{density matrix tau basis tensor} cancel in the tripartite information of three large adjacent blocks. The same conclusion is expected to hold quite generically for stationary states described by semilocal Gibbs ensembles. For arbitrary states the conclusion does not need to hold. For example, for states with non-negligible string contributions in \eqref{rdm as an average} {(in the $\tau$ representation)} the presented argument breaks right at the start. 

We note that the argument presented here is largely intuitive, despite being also quite technical. However, exact numerical results for the second R\'enyi entropy, obtained using fermionic techniques, provide an implicit check for the argument's validity.

\subsection{Jordan-Wigner transformation}\label{section jordan-wigner transformation}

The XY chain \eqref{eq:XY_model} is a quadratic form in Majorana fermions, defined as
\begin{equation}\label{majorana fermions}
    \bs a_{2\ell-1}\equiv \bs a^{x}_\ell=\left(\prod_{j<\ell}\bs\tau^z_j\right)\bs \tau^{x}_\ell , \qquad \bs a_{2\ell}\equiv \bs a^{y}_\ell=\left(\prod_{j<\ell}\bs\tau^z_j\right)\bs \tau^{y}_\ell \; .
\end{equation}
The Majorana fermions are self-adjoint and satisfy the algebra
\begin{equation}
    \{\bs a_{\ell}, \bs a_n\}=2\delta_{\ell n}\bs I \; .
\end{equation}
The XY chain Hamiltonian can due to translational invariance be written as
\begin{equation}
    \bs H = \frac{1}{4}\sum_{\ell,n=-\infty}^\infty\begin{pmatrix}\bs a^x_{\ell} & \bs a^y_{\ell}\end{pmatrix}
   \int_{-\pi}^{\pi}\tfrac{{\rm d}k}{2\pi}e^{i(\ell-n)k}\mathcal{H}(k)
    \begin{pmatrix}\bs a^x_{n} \\ \bs a^y_{n}\end{pmatrix},\label{eq:symbol_xy}
\end{equation}
where the $2\times 2$ matrix $\mathcal{H}(k)$, called the Hamiltonian symbol \cite{Fagotti2016Charges}, generates the couplings through its Fourrier coefficients. The symbol is given by
\begin{align}
\label{eq:symbol_H}
    \mathcal{H}(k)=2(J_x-J_y)\sin (k)\, \sigma^x-2(J_x+J_y)\cos(k)\, \sigma^y.
\end{align}
The positive eigenvalues of the symbol, given by
\begin{equation}    \varepsilon(k)=2\sqrt{J_x^2+J_y^2+2J_xJ_y\cos(2k)}
\end{equation}
are the energies of the quasiparticle excitations of the model.

In a translationally invariant state $\ket{\Psi}$ the correlation matrix
\begin{equation}
    \Gamma_{\ell,n}=\delta_{\ell,n}I-\bra{\Psi}\bs a_{\ell}\bs a_n\ket{\Psi}
\end{equation}
can be expressed in terms of its symbol $\Gamma(k)$, which is a $2\times 2$ matrix, as
\begin{equation}
\Gamma_{2\ell+i,2n+j}=  \int_{-\pi}^{\pi}\tfrac{{\rm d}k}{2\pi}e^{i(\ell-n)k}\Gamma_{ij}(k) \; ,\qquad \ell,n\in\mathbb{Z}, \ i,j\in\{1,2\}.
\end{equation}

In the dual picture the all-spin-up quench protocol is a homogenous quench protocol, while the flipped-spin protocol is a bipartitioning protocol. Accordingly, the correlation matrix symbol describing their stationary states has already been derived in the literature (see e.g. \cite{FagottiMaricZadnik2022,Maric2023Universality}). For the stationary state of the all-spin-up quench protocol it reads
\be\label{correlation matrix symbol GGE}
\textrm{protocol \ref{quench protocol 1}:}\qquad \Gamma(k)=\frac{\tr[\sigma^y\mathcal H(k)]\mathcal H(k)}{2\varepsilon^2(k)}\, ,
\ee
which can be obtained simply as a time average. The non-equilibrium stationary state following the quench from the domain wall state is also translationally invariant, notwithstanding that the initial state is not, and its correlation matrix symbol reads
\begin{equation}\label{correlation matrix symbol domain wall ness}
\textrm{protocol \ref{quench protocol 2}:}\qquad \Gamma(k)=-\mathrm{sgn}[v(k)]\frac{\tr[\sigma^y\mathcal H(k)]\mathbb I}{2\varepsilon(k)}\, ,
\end{equation}
where $v(k)=\varepsilon'(k)$ is the velocity of the quasiparticle excitations. The symbol \eqref{correlation matrix symbol domain wall ness} is derived using the generalized
hydrodynamic equation, which provides the connection between stationary states at different rays $\zeta=d/t$, where $d$ is the
distance from the position of the domain wall in the initial state.

The spin correlation functions of section \ref{section results spin correlations} and the string order parameters of section \ref{section results string order} are computed straightforwardly from the correlation matrix by expressing the operators in question in terms of fermions \eqref{majorana fermions}, where we use also some simplifying properties of the correlation matrix in question, as discussed in appendix \ref{appendix majorana correlators}. The R\'enyi entanglement entropy of a single block is obtained, as already done in \cite{FagottiMaricZadnik2022}, by expressing the density matrix \eqref{rdm connected block} in terms of fermionic Gaussians, i.e. exponentials of quadratic forms of fermionic operators. Similarly, the entanglement entropy of disjoint blocks is obtained by expressing the density matrix \eqref{rdm as an average} in terms of Gaussians. The procedure is pedantic, but the methods essentially do not go beyond ref. \cite{Fagotti2010disjoint} dealing with the entanglement entropy of disjoint blocks for the model \eqref{eq:XY_model}. The procedure of expressing the density matrices in terms of fermionic Gaussians is reported in appendix \ref{section rdm in the fermionic picture}, while the exact formulas for computing the second R\'enyi entropy are given in appendix \ref{section second Renyi entropy}.

\section{Conclusions}\label{section conclusions}

We have shown that the stationary state of the translationally invariant all-spin-up quench protocol and the stationary state reached in the flipped-spin quench protocol (NESS) have different properties. The former exhibits exponential decay of spatial correlations and zero tripartite information of three large adjacent subsystems, while in the latter there are correlations that decay only algebraically and the tripartite information is non-zero. On the other hand, string order is present in both cases. {It would be interesting to understand further the meaning of different string order parameters.} Importantly, a single spin flip in the initial state is responsible for differences in the behavior of the spatial correlations and the tripartite information between the two stationary states. This result complements the recents findings that a spin flip in the initial state affects the magnetization \cite{Fagotti2022Global} and the subleading term in the single block entanglement entropy \cite{FagottiMaricZadnik2022} in the stationary state. The mechanism responsible for all this phenomenology is the existence of semilocal charges, whose expectation value can be strongly affected by localized perturbations. We have also derived explicitly the expressions for generalized Gibbs ensembles describing the stationary state of the all-spin-up and the flipped-spin quench protocol. The set of semilocal charges associated with nonzero Lagrange multipliers is completely different for the two protocols.

We remark that if the initial state contains several flipped spins instead of just one the properties of the stationary state will strongly depend on the parity of the number of flipped spins. Flipping an even number of spins in the initial state is expected to yield a stationary state with the same properties as in the all-spin-up quench protocol, while an odd number of flipped spins is expected to yield the discovered phenomenology of the flipped-spin quench protocol. The reason is that flipping an even number of spins does not excite semilocal charges, while an odd number of flips does. 

The tripartite information in the stationary state of the flipped-spin quench protocol is interesting on its own. Although the initial state is a product state and the time evolution is with a gapped Hamiltonian, the tripartite information in the stationary state shares universal properties with conformal field theory. It depends on the lengths of the subsystems only through the cross ratio \eqref{eq:cross ratio definition}. It would be nice to have an explanation for this phenomenon independent of the details of the computation. {Let us note that, although the obtained formula for the tripartite information, given by eq. \eqref{tripartite domain wall any alpha}, is independent of the model parameters, such an independence would be lost in the non-interacting generalization of the model, the dual generalized XY chain introduced in ref. \cite{FagottiMaricZadnik2022}, for which the methods developed in this work still apply. Based on the results of Paper II, we conclude, on the other hand, that universality, i.e. the dependence on the configuration only through the cross ratio, should still hold. Another aspect of our result is that, as in protocols studied in Papers I and II,} the stationary state exhibits nonzero residual tripartite information. Namely, in the limit of small size of subsystem $B$ with respect to the size of subsystems $A,C$ (but still much larger than the lattice spacing) the tripartite information is nonzero {and equal to $-\log 2$}. This value should be contrasted to the zero value found in equilibrium at any temperature, irrespectively of criticality, or in other non-equilibrium settings, such as after quenches from ground states of gapped Hamiltonians. {It is an interesting open question whether universality would survive the generalization to interacting models and what would be the value of the residual tripartite information in this case.} 
Finally, we mention that the tripartite information is negative, which in other contexts \cite{Caceffo2023,Hayden2013Holographic} has been interpreted as an indication that quantum entanglement dominates over classical correlations. Maybe computing other entanglement measures, such as entanglement negativity \cite{Vidal2002,Plenio2005,Calabrese2002negativityPRL}, could shed light on the nature of this phase.

\section{Acknowledgments}
This work was supported by the European Research Council under the Starting Grant No. 805252 LoCoMacro. I thank Maurizio Fagotti for providing me the idea for this work, collaboration on related topics and useful discussions. I also thank Saverio Bocini for useful discussions.

\appendix

\section{Correlation functions of Majorana fermions}
\label{appendix majorana correlators}

Majorana correlation functions, determined by the symbols \eqref{correlation matrix symbol GGE} and \eqref{correlation matrix symbol domain wall ness}, have some simplifying properties. We assume $0<|J_y|<|J_x|$. In the stationary state of quench protocol \ref{quench protocol 1} the correlations are
\begin{equation}
\begin{split}
     &\braket{\bs a_{\ell+r}^x\bs a_{\ell}^y}\\
    &=\begin{cases}
    0, & r<0 \textrm{ or } r \in 2\mathbbm{Z} +1\\
    \frac{i}{2}\big(1+\frac{J_y}{J_x}\big), & r=0 \\
\frac{i}{2}\left(1-\frac{J_y^2}{J_x^2}\right)\left(-\frac{J_y}{J_x}\right)^{\frac{r}{2}-1}, & r>0 \textrm{ and } r\in 2\mathbb{Z}
    \end{cases} \; ,
\end{split}
\end{equation}
as obtained by expanding \eqref{correlation matrix symbol GGE} in the Fourier series, where it is convenient to exploit the formula for the geometric series. The correlations vanish for all odd $r$ and for negative even $r$. Moreover the correlations decay exponentially with $r$, related to the fact that symbol \eqref{correlation matrix symbol GGE} is smooth. Also, trivially, the correlations $\braket{\bs a_{\ell+r}^\alpha\bs a_{\ell}^\beta}$ for $\alpha=\beta$ vanish for all $r\neq 0$. 

In the stationary state of quench protocol \ref{quench protocol 2} the correlations $\braket{\bs a_{\ell+r}^\alpha\bs a_{\ell}^\beta}$ vanish for $\alpha\neq \beta$, while for $\alpha=\beta$ they vanish for nonzero even $r$. The remaining ones decay to zero algebraically with $r$, since \eqref{correlation matrix symbol domain wall ness} is discontinuous. Using partial integration we get
\begin{equation}
    \braket{\bs a_{\ell+2r-1}^x \bs a_{\ell}^x}=
    \braket{\bs a_{\ell+2r-1}^y \bs a_{\ell}^y}
    \simeq-\mathrm{sgn}(J_y) \frac{i}{\pi r}.
\end{equation}
In computing the string order parameter we will also use the exact result for $r=1$
\begin{equation}
   \braket{\bs a_{\ell+1}^x  \bs a_{\ell}^x}=  \braket{\bs a_{\ell+1}^y \bs a_{\ell}^y}=-\mathrm{sgn}(J_y)\frac{i}{\pi}\left(1+\frac{J_y}{J_x}\right) \; ,
\end{equation}
obtained by basic integral manipulations.

The spin correlation functions in section \ref{section results spin correlations} and the string order parameters in section \ref{section results string order} are computed straightforwardly using the Kramers-Wannier transformation \eqref{eq:dualTDeven} and the listed properties of Majorana correlators. In particular, the string operators read
\begin{align}
  &  \prod_{\ell=-r}^r\bs\sigma^z_\ell =\bs \tau^z_{-r}\bs \tau^z_{r+1}=-\bs a_{-r}^x \bs a_{-r}^y \bs a_{r+1}^x\bs a_{r+1}^y \; , \\
  & \bs\sigma_{-r-3}^x\bs\sigma_{-r-2}^z\bs\sigma_{-r-1}^y\left(\prod_{\ell=-r}^{r}\bs\sigma^z_\ell\right)\bs\sigma^y_{r+1}\bs\sigma_{r+2}^z\bs\sigma_{r+3}^x=\bs\tau_{-r-2}^y\bs\tau_{-r-1}^x\bs\tau_{r+2}^x\bs\tau_{r+3}^y=\bs a_{-r-2}^x \bs a_{-r-1}^x \bs a_{r+2}^y \bs a_{r+3}^y \; .
\end{align}

\section{Procedure for finding the Lagrange multipliers}\label{appendix Lagrange multipliers}

In this section we show how the Lagrange multipliers in the generalized Gibbs ensembles describing the studied stationary states are obtained, based on methods developed in \cite{Fagotti2013Reduced,Fagotti2014On}.

The one-site shift invariant charges of the quantum XY chain \eqref{eq:XY_model} given in section \ref{section semilocal charges} can be written as a quadratic form in Majorana fermions
\begin{equation}
    \bs Q^{(n,\pm)} = \frac{1}{4}\sum_{\ell,n=-\infty}^\infty\begin{pmatrix}\bs a^x_{\ell} & \bs a^y_{\ell}\end{pmatrix}
   \int_{-\pi}^{\pi}\tfrac{{\rm d}k}{2\pi}e^{i(\ell-n)k}\mathcal{Q}^{(n,\pm)}(k)
    \begin{pmatrix}\bs a^x_{n} \\ \bs a^y_{n}\end{pmatrix},\label{eq:symbol charge}
\end{equation}
specified by the symbol $\mathcal{Q}^{(n,\pm)}(k)$, which is a $2\times 2$ matrix function. The charges are constructed so that the symbol $\mathcal{Q}^{(n,\pm)}(k)$ commutes with the Hamiltoian symbol $\mathcal{H}(k)$, because this implies that charges commute with the Hamiltonian ($[\bs Q^{(n,\pm)}, \bs H]=0$). Moreover, the symbol is chosen to have a finite number of Fourier coefficients, to ensure the locality of the charge. In particular, the charges of section \ref{section semilocal charges} are given by the symbols
\begin{align}
   & \mathcal{Q}^{(n,+)}(k)=\cos(nk)\mathcal{H}(k) \; ,\\
   & \mathcal{Q}^{(n,-)}(k)=\sin[(n+1)k]\mathbb{I} \; ,
\end{align}
for $n\in\mathbb{N}_0$.

A translationally invariant fermionic Gaussian can be written as
\begin{equation}
    \bs\rho=\frac{1}{\mathcal{Z}}\exp \left[\frac{1}{4}\sum_{\ell,n=-\infty}^\infty\begin{pmatrix}\bs a^x_{\ell} & \bs a^y_{\ell}\end{pmatrix}
   \int_{-\pi}^{\pi}\frac{{\rm d}k}{2\pi}e^{i(\ell-n)k}\mathcal{W}(k)
    \begin{pmatrix}\bs a^x_{n} \\ \bs a^y_{n}\end{pmatrix}\right],
\end{equation}
where $\mathcal{Z}$ is the normalization and $\mathcal{W}(k)$ is related to the symbol of the correlation matrix (see e.g. \eqref{correlation matrix Gamma W relation}) through
\begin{equation}\label{Gamma W symbol relation}
    \mathcal{W}(k)=2\ \mathrm{artanh}\left[\Gamma(k)\right] \;.
\end{equation}
On the other hand, $\mathcal{W}(k)$ is related to the Lagrange multipliers in the GGE \eqref{GGE XY general}
through
\begin{equation}\label{symbol W Q relation}
    \mathcal{W}(k)=-\sum_{n=0}^\infty \left(\lambda^{(n,+)}\mathcal{Q}^{(n,+)}(k)+\lambda^{(n,-)}\mathcal{Q}^{(n,-)}(k)\right) \; .
\end{equation}
Thus, Lagrange multipliers $\{\lambda^{(n,+)},\lambda^{(n,-)}\}_n$ are obtained from the correlation matrix symbol, given by \eqref{correlation matrix symbol GGE} and \eqref{correlation matrix symbol domain wall ness} for the studied quench protocols, by comparing \eqref{Gamma W symbol relation} and \eqref{symbol W Q relation}. In this way the Lagrange multipliers in \eqref{lagrange multiplier protocol 1} and \eqref{lagrange multiplier protocol 2} are obtained, where the given integrals are simplifications of the ones from $0$ to $2\pi$ arising from the Fourier coefficients.

\section{Reduced density matrix in the fermionic picture}\label{section rdm in the fermionic picture}
The usefulness of representation \eqref{rdm connected block} and \eqref{rdm as an average} is that they enable us to go to the fermionic language. The goal of this section is to use these representations to express the density matrices in a useful form for computing the R\'enyi entropies, which is the subject of appendix \ref{section second Renyi entropy}. Assuming $\min A =1$ without loss of generality, the mapping starts by defining the Majorana fermions restricted to $\tilde A\tilde B\tilde C$,
\begin{equation}\label{eq:MajoranaRestricted}
   a_{2\ell-1}\equiv  a^{x}_\ell=\left(\prod_{j=1}^{\ell-1} \tau^z_j\right) \tau^{x}_\ell , \qquad  a_{2\ell}\equiv  a^{y}_\ell=\left(\prod_{j=1}^{\ell-1} \tau^z_j\right) \tau^{y}_\ell \; ,
\end{equation}
for $\ell=1,2,\ldots , |\tilde A\tilde B\tilde C|$, which are analogous to those defined on the whole chain and satisfy analogous algebra $\{ a_{\ell}, a_n\}=2\delta_{\ell n} I$.

\paragraph{Gaussians of fermionic operators} In the $2^d$ dimensional Fock space associated to Majorana fermions $a_1,a_2,a_3,\ldots ,a_{2d}$ (with anticommutation relations $\{a_{\ell},a_n\}=2\delta_{\ell,n}$) a generic normalized Gaussian operator $\rho(\Gamma)$ reads
\begin{equation}\label{gaussian}
	\rho(\Gamma)=\frac{1}{\mathcal{Z}}\exp\left(\frac{\vec{a}^\dagger W \vec{ a}}{4}\right)\,,
\end{equation}
where $\vec{a}^\dagger=(a_1,a_2,\ldots ,a_{2d})$ and $W$ is a $2d\times 2d$ dimensional antisymmetric matrix with complex entries. The Gaussian is completely specified by its two-point correlation matrix, given by
\begin{equation}\label{correlation matrix definition appendix}
\Gamma_{j,\ell}=\delta_{j,\ell}-\tr \left[ \rho(\Gamma) \
 a_j a_{\ell} \right],\qquad  1\leq j,l \leq 2d ,
\end{equation}
which is related to matrix $W$ in the exponent through the formula \cite{Fagotti2010disjoint}
\begin{equation}\label{correlation matrix Gamma W relation}
    \Gamma=\tanh \left(\frac{W}{2}\right).
\end{equation}
The normalization is given by
\begin{equation}\label{normalization}
	\mathcal{Z}=\pm \left[\det\left(e^{\frac{W}{2}}+e^{-\frac{W}{2}}\right)\right]^{\frac{1}{2}} = \pm \left[\det\left(\frac{I-\Gamma}{2}\right)\right]^{-\frac{1}{2}},
\end{equation}
where the sign in front of the square root is not specified by the formula and depends on further details of $\Gamma$. When $\rho$ is a density matrix we have that $\Gamma$ and $W$ are Hermitian so the sign is positive. We will encounter non-Hermitian $\Gamma$ and $W$ only in the case of disjoint blocks, analogously to ref. \cite{Fagotti2010disjoint}.

\paragraph{Single block} In the case of a single block, it is an established fact \cite{Peschel2003,Peschel2009,Vidal2003,Latorre2004} that the reduced density matrices such as $\tilde\rho_{\tilde A}$ in \eqref{rdm connected block}, defined by \eqref{eq:rdm rho A}, are Gaussians, i.e. exponentials of quadratic forms of Majorana fermions, determined by the correlation matrix. As discussed in \cite{FagottiMaricZadnik2022} and in appendix \ref{section second Renyi entropy}, each term in the sum in \eqref{rdm connected block} is still a Gaussian, with the correlation matrix depending on the transformations $P$. The Gaussian structure can then be exploited for the computation of the R\'enyi entropies.

\paragraph{Disjoint blocks} In the case of disjoint blocks we have derived representation \eqref{rdm as an average} for the reduced density matrix. Each term in the sum in \eqref{rdm as an average} can be written as a sum of Gaussians. The Gaussian structure can then be exploited to compute the Rényi entropies.

Expressing $\tilde \rho_{\tilde A\tilde C}$, appearing in \eqref{rdm as an average}, in terms of Gaussians is a procedure identical to expressing the reduced density matrix in the XY chain in terms of Gaussians, as done in ref. \cite{Fagotti2010disjoint}. We review it briefly because we use the same techniques for $ \omega_{\tilde A\tilde C}$. 
We have
\begin{equation}
\begin{split}\label{step rho fermion basis sum}
    \tilde \rho_{\tilde A\tilde C} &=\frac{1}{2^{|\tilde A \tilde B \tilde C|}}\sum_{\mathrm{even}\;F_{\tilde A}, F_{\tilde C}}\bra{\Psi}F_{\tilde A} F_{\tilde C}\ket{\Psi} \left( F_{\tilde A} F_{\tilde C}\right)^\dagger\\&+\frac{1}{2^{|\tilde A \tilde B \tilde C|}}\sum_{\mathrm{odd}\;F_{\tilde A}, F_{\tilde C}}\bra{\Psi}F_{\tilde A} F_{\tilde C}S^z\ket{\Psi} \left( F_{\tilde A} F_{\tilde C}S^z\right)^\dagger\; , \\
    \end{split}
\end{equation}
where the first (second) sum is over all $F_{\tilde A}$ and $F_{\tilde C}$ that are both a product of an even (odd) number fermions in $\tilde A$ and $\tilde C$ respectively, i.e. $F_{\tilde A}$ and $F_{\tilde C}$ in the first (second) sum commutes (anticommutes) with $\Pi_\tau^z\equiv\prod_{\ell\in A}\tau_\ell^z$.
Explicitly,
\begin{equation}
         F_{\tilde A}=\prod_{\ell\in \tilde A}(a_\ell^x)^{j^x_\ell}(a_\ell^y)^{j^y_\ell} ,\qquad  F_{\tilde C}=\prod_{\ell\in \tilde C}(a_\ell^x)^{j^x_\ell}(a_\ell^y)^{j^y_\ell} \; ,
\end{equation}
where indices $j_k^{x,y}$ are either $0$ or $1$ and the order of operators in the product does not matter in \eqref{step rho fermion basis sum}. In the second term in \eqref{step rho fermion basis sum} the string $S^z$, where we denote $S^\gamma=\prod_{\ell\in \tilde B}\tau^\gamma_\ell$ for $\gamma=0,x,y,z$, appears in order to cancel the string that arises due to the non-locality of the Jordan-Wigner transformation. The two different sums can be expressed conveniently using the (anti)commutation relation with the string over the block $\tilde A$
\begin{equation}
    P_0\equiv S_{\tilde A}^z= \prod_{\ell\in \tilde A} \tau_\ell^z \; .
\end{equation}
We have
\begin{equation}\label{eq:rho 1 four gaussians}
    \tilde \rho_{\tilde A \tilde C}=\frac{\rho_0+S_{\tilde A}^z\rho_0S_{\tilde A}^z}{2}+S^z\frac{\rho_z-S_{\tilde A}^z\rho_zS_{\tilde A}^z}{2},
\end{equation}
where we denote
\begin{equation}\label{rho alpha definition}
  \rho_\gamma=\frac{1}{2^{|\tilde A \tilde B\tilde C|}}  \sum_{F_{\tilde A}, F_{\tilde C}}\bra{\Psi}F_{\tilde A} F_{\tilde C}S^\gamma\ket{\Psi} \left( F_{\tilde A} F_{\tilde C}\right)^\dagger
\end{equation}
for $\gamma=0,x,y,z$. Here the sum is over all possible products of fermions in $\tilde A$ and $\tilde C$.

The operator $\omega_{\tilde A\tilde C}$, appearing in \eqref{rdm as an average}, can be expressed in terms of fermions in a similar way. For even $|\tilde B|$ it is completely analogous to $\tilde \rho_{\tilde A\tilde C}$,
\begin{equation}
\begin{split}
    \omega_{\tilde A\tilde C} &=S^x\frac{1}{2^{|\tilde A \tilde B \tilde C|}}\sum_{\mathrm{even}\;F_{\tilde A}, F_{\tilde C}}\bra{\Psi}F_{\tilde A} F_{\tilde C}S^x\ket{\Psi} \left( F_{\tilde A} F_{\tilde C}\right)^\dagger\\&+S^x\frac{1}{2^{|\tilde A \tilde B \tilde C|}}\sum_{\mathrm{odd}\;F_{\tilde A}, F_{\tilde C}}\bra{\Psi}F_{\tilde A} F_{\tilde C}S^zS^x\ket{\Psi} \left( F_{\tilde A} F_{\tilde C}S^z\right)^\dagger\; , \\
    \end{split}
\end{equation}
where the first (second) sum is again over all $F_{\tilde A}$ and $ F_{\tilde C}$ that are both a product of an even (odd) number of fermions and $S^z$ stems from the non-locality of the Jordan-Wigner transformation. For odd $|\tilde B|$ it is slightly different,
\begin{equation}
\begin{split}
   \omega_{\tilde A\tilde C} & =S^x\frac{1}{2^{|\tilde A \tilde B \tilde C|}}\sum_{\mathrm{odd}\;F_{\tilde A},\; \mathrm{even}\;F_{\tilde C}}\bra{\Psi}F_{\tilde A} F_{\tilde C}S^x\ket{\Psi} \left( F_{\tilde A} F_{\tilde C}\right)^\dagger\\&+S^x\frac{1}{2^{|\tilde A \tilde B \tilde C|}}\sum_{\mathrm{even}\;F_{\tilde A},\; \mathrm{odd}\;F_{\tilde C}}\bra{\Psi}F_{\tilde A} F_{\tilde C}S^zS^x\ket{\Psi} \left( F_{\tilde A} F_{\tilde C}S^z\right)^\dagger\; . \\
    \end{split}
\end{equation}
Now the number of fermions in $F_{\tilde A}$ and $ F_{\tilde C}$ has opposite parities. The string $S^z$ has to be included when $F_{\tilde C}$ consists of an odd number of fermions or, consequently, when $F_{\tilde A}$ consists of an even number of fermions.

Overall, we have
\begin{equation}\label{eq:rho 2 four gaussians}
     \omega_{\tilde A \tilde C}=S^x \frac{\rho_x+(-1)^{|\tilde B|}S_{\tilde A}^z\rho_x S_{\tilde A}^z}{2}+S^y\frac{\rho_y- (-1)^{|\tilde B|}S_{\tilde A}^z\rho_y S_{\tilde A}^z}{2},
\end{equation}
where the string $S^y$ is a result of multiplying $S^x$, that comes from the non-local character of the Kramers-Wannier transformation, with $S^z$, that comes from the non-locality of the Jordan-Wigner transformation. Finally, \eqref{eq:rho 2 four gaussians} can be expressed in terms of Gaussians, since the operators \eqref{rho alpha definition} can be expressed in terms of Gaussians, as covered in the following.

\paragraph{Operators with strings as Gaussians}
Let us consider operators defined in \eqref{rho alpha definition}. We note that they satisfy
    \begin{equation}\label{new string first appearance}
  S^\gamma\rho_\gamma=\frac{1}{2^{|\tilde A \tilde B\tilde C|}}  \sum_{F_{\tilde A}, F_{\tilde C}}\bra{\Psi}F_{\tilde A} F_{\tilde C}\mathcal{S}^\gamma\ket{\Psi} \left( F_{\tilde A} F_{\tilde C}\mathcal{S}^\gamma\right)^\dagger,
\end{equation}
where we have the freedom of replacing the string between the blocks $S^\gamma$ by $\mathcal{S}^\gamma=F'_{\tilde A}F'_{\tilde C}S^\gamma$ for any fixed $F_{\tilde A}'$ and $F_{\tilde C}'$, that are a product of fermionic operators in $\tilde A$ and $\tilde C$ respectively. The exact choice is discussed in appendix \ref{appendix strings in the reduced density matrix of disjoint blocks}. This replacement is allowed since it corresponds simply to changing the dummy summation index. Because of this freedom we can choose $\mathcal{S}^\gamma$ to be a product of an even number of Majorana fermions (irrespectively of the parity of $|\tilde B|$) and therefore a Gaussian. Moreover, choosing $\mathcal{S}^\gamma$ such that it has a non-zero expectation value $\braket{\mathcal S^\gamma}$, the expectation values
\begin{equation}\label{step constructing new string}
\braket{F_{\tilde A}F_{\tilde C}\mathcal{S}^\gamma}=\braket{\mathcal{S}^\gamma}\tr\left[F_{\tilde A}F_{\tilde C} \rho(\Gamma_\gamma)\right],   
\end{equation}
are given by the Gaussian $\rho(\Gamma_\gamma)$ specified by the correlation matrix
\begin{equation}\label{eq:correlation matrix string}  \left(\Gamma_\gamma\right)_{j,\ell}=\delta_{j\ell}-\frac{\braket{a_{f(j)}a_{f(\ell)} \mathcal{S}^\gamma}}{\braket{\mathcal{S^\gamma}}}\; ,\qquad j,\ell=1,2,\ldots,2|\tilde A\tilde C|
\end{equation}
where
\begin{equation}
    f(j)=\begin{cases}
       &j , \qquad 1\leq j\leq |2\tilde A|\\
       &2[\min{(\tilde C)}-|\tilde A|]+j -2 , \qquad 2|\tilde A|+1\leq j\leq 2|\tilde A\tilde C|
    \end{cases}
\end{equation}
is just a compact notation to work with the indices related to disjoint blocks.
Here we have used the Wick theorem, the property that the product of two Gaussians ($\mathcal{S}^\gamma$ and $\rho(\Gamma)$) is a Gaussian \cite{Fagotti2010disjoint} and we have divided by the expectation value of the string $\braket{\mathcal{S}^\gamma}$ to ensure the normalization $\tr \left[\rho(\Gamma_\gamma)\right]=1$. The correlation matrix \eqref{eq:correlation matrix string} and, accordingly, the matrix $W$ in \eqref{gaussian}, are Hermitian for $\gamma=0$, but in general this is not the case. In general they are complex antisymmetric matrices. We note that Wick theorem holds also in this more general case. It is desirable to have a proof of Wick theorem as such directly in the Majorana fermions formalism we work with so we provide it in appendix \ref{appendix wick theorem}.

We have thus shown that $S^\gamma\rho_\gamma$ can be expressed in terms of a normalized Gaussian,
\begin{equation}
S^\gamma\rho_\gamma=\left(\mathcal{S}^\gamma\right)^\dagger\braket{\mathcal{S}^\gamma} \rho(\Gamma_\gamma)\; .
\end{equation}
The correlation matrix \eqref{eq:correlation matrix string} could, in principle, be evaluated from its definition, but since this procedure would require the evaluation of the pfaffian for each matrix element, in appendix \ref{appendix correlation matrix with strings} we derive an alternative expression (a similar trick was used in \cite{Fagotti2010disjoint} to avoid singularities). Suppose that the string is given by $\mathcal{S}^\gamma=a_{i_1}a_{i_2}\ldots a_{i_m}$ for some positive even integer $m$. We find
\begin{equation}\label{eq:correlation matrix string workable expression}
    \Gamma_\gamma=\Gamma_0-UN^{-1}U^{\mathrm{T}},
\end{equation}
where
\begin{equation}\label{matrix U definition rectangular}
    U_{\ell,n}=\braket{a_{f(\ell)}a_{i_n}} \qquad \ell =1,2,\ldots, 2|\tilde A \tilde C|, \; n=1,2,\ldots m
\end{equation}
and
\begin{equation}\label{matrix N definition string}
N_{\ell,n}=\braket{a_{i_\ell}a_{i_n}} \; , \qquad \ell,n =1,2,\ldots,m \; .
\end{equation}
Note that the squared absolute value of the expectation value of the string is given by
\begin{equation}
    |\braket{\mathcal{S}^\gamma}|^2=|\det N|. 
\end{equation}

\section{The second Rényi entropy}\label{section second Renyi entropy}
In the previous section we have derived fermionic representations of reduced density matrices. In this section we use these expressions to find a way for computing the R\'enyi entropies exactly. For simplicity, we focus only on the second Rényi entropy.

\paragraph{Products of Gaussians} First we cover some basic formulas. For the computation of the second Rényi entropy we only need the knowledge of the trace of the product of two normalized Gaussians, defined in \eqref{gaussian}. It has been argued in \cite{Fagotti2010disjoint} that for two normalized Gaussians $\rho(\Gamma_1)$ and $\rho(\Gamma_2)$, with correlation matrices $\Gamma_1$ and $\Gamma_2$, the trace is given by
\begin{equation}\label{trace of a product of gaussians}
    \tr\left[\rho(\Gamma_1)\rho(\Gamma_2)\right]\equiv\mathcal{G}(\Gamma_1,\Gamma_2)\equiv\prod_{\{\lambda\}/2}\lambda \; ,
\end{equation}
where the product is over all eigenvalues $\lambda$ of the matrix $(I+\Gamma_1\Gamma_2)/2$ with degeneracy, that is always even, reduced by half. Note that up to an unresolved sign the following formula holds (see also \cite{Klich2014})
\begin{equation}\label{trace of a product of gaussians unresolved sign}
    \mathcal{G}(\Gamma_1,\Gamma_2)=\pm\sqrt{\det\left(\frac{I+\Gamma_1\Gamma_2}{2}\right)} \; .
\end{equation}

\paragraph{Single block}
The expression for computing exactly the second R\'enyi entropy of a single block has already been derived in \cite{FagottiMaricZadnik2022}. We revisit it here since we need the result for computing the tripartite information, and as a warm-up for the case of disjoint blocks. In the case of a single block the density matrix $\tilde \rho_{\tilde A}$ in \eqref{rdm connected block} is a Gaussian,
\begin{equation}
    \tilde \rho_{\tilde A} =\rho(\Gamma)
\end{equation}
where the correlation matrix is given by $\Gamma=\delta_{\ell,n}-\braket{a_\ell a_n}$ for $\ell,n=1,2,\ldots,2|\tilde A|$. The transformations $P_j$ in \eqref{rdm connected block} preserve the Gaussianity of the state but modify the correlation matrix. The latter can be found simply by studying the effects on the two-point products of Majorana fermions. Namely, the operator $P_j\rho(\Gamma)P_j$, for $j=1,2,3$, is the (normalized) Gaussian $\rho(V_j\Gamma V_j)$, where $V_j$ are diagonal matrices of size $2|\tilde A|$ defined by
\begin{equation}
\left(V_{1}\right)_{\ell\ell}=\begin{cases}
        -1 , & \ell=1,2\\
        1, & \mathrm{otherwise} 
    \end{cases} \; , 
    \qquad
   \left(V_{2}\right)_{\ell\ell}=\begin{cases}
        -1 , & \ell=2|\tilde A|-1, 2|\tilde A|\\
        1, & \mathrm{otherwise} 
    \end{cases} \; , 
    \qquad
    \left(V_3\right)_{\ell\ell}=(-1)^{\lfloor \frac{\ell}{2}\rfloor} \; ,
\end{equation}
where $\lfloor.\rfloor$ stands in the standard way for the floor function.
The R\'enyi entropy now follows by applying the formula \eqref{trace of a product of gaussians} for the trace of a product of two Gaussians. Introducing the compact notation
\begin{equation}
    \mathcal{V}[j_1,j_2,j_3](\Gamma)=\left(V_{1}^{j_1}V_{2}^{j_2}V_3^{j_3}\right)^{\mathrm{T}}\Gamma \left(V_{1}^{j_1}V_{2}^{j_2}V_3^{j_3}\right) \;,
\end{equation}
we have
\begin{equation}
    S_2(A)=2\log 2-\log \left[\sum_{j_1,j_2,j_3=0}^1\mathcal{G}\Big(\Gamma,\mathcal{V}[j_1,j_2,j_3](\Gamma)\Big)\right],
\end{equation}
where $\mathcal{G}$ is given by \eqref{trace of a product of gaussians unresolved sign} with the $+$ sign. Note that the summand $2\log 2$ comes from the second term in \eqref{entropy enlargement of the space} (for $|X'|=1$) and from the factor $1/2^3$ in \eqref{rdm connected block}.

\paragraph{Disjoint blocks}
The second R\'enyi entropy of disjoint blocks can be computed using eq. \eqref{rdm as an average}, relating the $\sigma$ and $\tau$ representation, together with eq. \eqref{eq:rho 1 four gaussians}, \eqref{eq:rho 2 four gaussians}, that allow us to express the density matrix as a sum of Gaussian fermionic operators. In order to cancel the string operators without additional phase factors, using $\left(\mathcal{S}^\gamma\right)^\dagger\mathcal{S}^\gamma=1$, we are going to start from the trivial relation $\tr(\bar{\rho}_{AC}^2)=\tr(\bar \rho_{AC} \bar \rho_{AC}^\dagger)$. Using the property
\begin{equation}
    \tr\left(\frac{\rho_1\pm P\rho_1P}{2}\frac{\rho_2\pm P\rho_2P}{2}\right)=\tr \left(\rho_1 \frac{\rho_2\pm P\rho_2P}{2}\right),
\end{equation}
that holds for any operators $\rho_1,\rho_2$ given some Hermitian involution $P$, we also have the simplification
\begin{equation}
    \tr\left(\bar \rho_{AC}^2\right)=\tr\left( \tilde \rho_{\tilde A\tilde C}\bar\rho_{AC}^\dagger+\omega_{\tilde A\tilde C}\bar\rho_{AC}^\dagger\right)
\end{equation}

It is crucial that each of the four terms appearing in \eqref{eq:rho 1 four gaussians} and \eqref{eq:rho 2 four gaussians} multiplied by the string is a Gaussian. Similarly to the single block case, the information on the effects of the transformations $S_{\tilde{A}}^z$ and $P_j$, for $j=1,2,\ldots,6$, appearing in \eqref{rdm as an average}, on Gaussian operators $\rho_\gamma$, for $\gamma=0,x,y,z$, are encoded in the effects on the correlation functions. Namely, for a normalized Gaussian $\rho(\Gamma)$ the operator $P_j\rho(\Gamma)P_j$ is for $j=0,1,\ldots, 6$ a normalized Gaussian $\rho(V_j\Gamma V_j)$, where $V_j$ are diagonal matrices of size $2|\tilde A\tilde C|$ defined by
\begin{align}
    & \left(V_0\right)_{\ell\ell}=\begin{cases}
        -1 , & \ell \leq 2|\tilde A|\\
        1, & \mathrm{otherwise}
    \end{cases} \; ,
\qquad    \left(V_3\right)_{\ell\ell}=\begin{cases}
        (-1)^{\lfloor \frac{\ell}{2}\rfloor} , & \ell \leq 2|\tilde A|\\
        (-1)^{|\tilde A|}, & \mathrm{otherwise}
    \end{cases}\; ,
      \qquad \left(V_6\right)_{\ell\ell}=\begin{cases}
        1 , & \ell \leq 2|\tilde A|\\
        (-1)^{\lfloor \frac{\ell}{2}\rfloor}, & \mathrm{otherwise}
    \end{cases} \; ,\\
    &\left(V_{j}\right)_{\ell\ell}=\begin{cases}
        -1 , & \ell=2r_j-1,2r_j\\
        1, & \mathrm{otherwise} 
    \end{cases} \qquad \textrm{for $j=1,2,4,5$, where $r_1=1,\ r_2=|\tilde A|, \ r_4=|\tilde A|+1 , \ r_5=|\tilde A \tilde C|$} \; .
\end{align}
To write the final formula for the second Rényi entropy let us introduce the signs $s_j^\gamma$, for $j=0,1,\ldots,6$, which are positive or negative depending on whether $P_j$ commutes or anticommutes with $\mathcal{S}^\gamma$ respectively, i.e.
\begin{equation}
    s_j^\gamma\equiv\begin{cases}
        1, & [P_j,\mathcal{S}^\gamma]=0\\
        -1,  & \{P_j,\mathcal{S}^\gamma\}=0
    \end{cases}
\end{equation}
We remind that the string operators $\mathcal{S^\gamma}$ are chosen so that $\braket{\mathcal{S}^\gamma}\neq 0$ (see appendix \ref{appendix strings in the reduced density matrix of disjoint blocks} for the exact choice). Note that if $\mathcal{S}^\gamma=S^\gamma$ all the signs $s_j^\gamma$ are positive.
Let us also introduce the compact notation
\begin{equation}
    s^\gamma_{j_1,j_2,\ldots ,j_6}=(s_1^\gamma)^{j_1}(s_2^\gamma)^{j_2}\ldots (s_6^\gamma)^{j_6} \; ,
\end{equation}
and, similarly to the single block case,
\begin{equation}
    \mathcal{V}[j_0,j_1,\ldots,j_6](\Gamma)=\left(V_{0}^{j_0}V_{1}^{j_1}\ldots V_6^{j_6}\right)^{\mathrm{T}}\Gamma \left(V_{0}^{j_0}V_{1}^{j_1}\ldots V_6^{j_6}\right) \; .
\end{equation}

The formula for the second R\'enyi entropy then reads
\begin{equation}\label{second renyi entropy final exact}
\begin{split}
       &S_2(AC)=5\log 2-\log\bigg\{ \sum_{j_{1,2,\ldots,6}=0}^1 \bigg(\\ 
       &\mathcal{G}\Big(\Gamma_0,\mathcal{V}[0,j_1,\ldots,j_6](\Gamma_0)\Big)+\mathcal{G}\Big(\Gamma_0,\mathcal{V}[1,j_1,\ldots,j_6](\Gamma_0)\Big)\\
    & +(-1)^{j_4+j_5} s^x_{j_1,j_2,\ldots,j_6}|\braket{\mathcal{S}^x}|^2 \left[\mathcal{G}\Big(\Gamma_x,\mathcal{V}[0,j_1,\ldots,j_6](\Gamma_x^\dagger)\Big)+s_0^x(-1)^{|\tilde B|}\mathcal{G}\Big(\Gamma_x,\mathcal{V}[1,j_1,\ldots,j_6](\Gamma_x^\dagger)\Big)\right]\\
    & +(-1)^{j_4+j_5} s^y_{j_1,j_2,\ldots,j_6}|\braket{\mathcal{S}^y}|^2 \left[\mathcal{G}\Big(\Gamma_y,\mathcal{V}[0,j_1,\ldots,j_6](\Gamma_y^\dagger)\Big)-s_0^y(-1)^{|\tilde B|}\mathcal{G}\Big(\Gamma_y,\mathcal{V}[1,j_1,\ldots,j_6](\Gamma_y^\dagger)\Big)\right]\\
    & + s^z_{j_1,j_2,\ldots,j_6}|\braket{\mathcal{S}^z}|^2\left[ \mathcal{G}\Big(\Gamma_z,\mathcal{V}[0,j_1,\ldots,j_6](\Gamma_z^\dagger)\Big)-s_0^z\mathcal{G}\Big(\Gamma_z,\mathcal{V}[1,j_1,\ldots,j_6](\Gamma_z^\dagger)\Big)\right]  \; \bigg)\bigg\},
\end{split}
\end{equation}
Note that the summand $5\log 2$ comes from second term in \eqref{entropy enlargement of the space}, the factor $1/2^{6}$ in \eqref{rdm as an average}, the factor $1/2$ in \eqref{eq:rho 1 four gaussians} and \eqref{eq:rho 2 four gaussians} and by taking into account that in applying the formula \eqref{trace of a product of gaussians} a factor $2^{|\tilde B|}$ has to be included to account for the degeneracy in the enlarged space we work with.

The final expression \eqref{second renyi entropy final exact} is rather lengthy. Evaluating the second R\'enyi entropy exactly requires the evaluation of 512 traces of products of Gaussians. In practical implementations we will notice that some terms become negligible with respect to the others, beyond the machine precision, very fast as we increase the size of the subsystem. To speed up our numerical implementation, once some terms become negligible as we increase the subsystem size, we simply drop them for larger subsystems.

\section{Choice of the strings for the reduced density matrix of disjoint blocks}\label{appendix strings in the reduced density matrix of disjoint blocks}

In appendices \ref{section rdm in the fermionic picture} and \ref{section second Renyi entropy} we have not specified the choice of the string operators $\mathcal{S}^\gamma$, $\gamma=x,y,z$, introduced in \eqref{new string first appearance} and appearing in the final formula \eqref{second renyi entropy final exact} for the second R\'enyi entropy. As discussed after eq. \eqref{new string first appearance}, the string operators are given by
\begin{equation}
    \mathcal{S}^\gamma=F_{\tilde A}'F_{\tilde C}'S^\gamma
\end{equation}
where $S^\gamma=\prod_{\ell\in \tilde B}\tau^\gamma_\ell$ and $F_{\tilde A}', F_{\tilde C}'$ are products of fermions whose indices are restricted to block $\tilde A, \tilde C$ respectively. The operators $F_{\tilde A}', F_{\tilde C}'$ have to be chosen in such a way that $\braket{\mathcal S^\gamma}\neq 0$. We do this task by exploiting the properties of the Majorana correlators given in appendix \ref{appendix majorana correlators}.
In the stationary state of the flipped-spin quench protocol the string operators 
\begin{align}
&\mathcal{S}^\gamma=\left(\prod_{\ell=|\tilde A|-|\tilde B|}^{|\tilde A|-1}\sigma_{\ell}^\gamma\right) S^\gamma, \qquad \gamma=x,y\\
&\mathcal{S}^z=\begin{cases}
    S^z, & |\tilde B| \textrm{ even}\\
    \tau^z_{|\tilde A|}S^z, & |\tilde B| \textrm{ odd}
\end{cases}
\end{align}
do the job when $|\tilde A|>|\tilde B|$. For $|\tilde A|\leq|\tilde B|$ in the studied examples (the uppermost curve in figure \ref{fig:xdependence}) the same $\mathcal{S}^z$ works, while we have been able to show directly that all correlation functions of the form $\braket{F_{\tilde A}F_{\tilde C}S^\gamma}$ vanish for $\gamma=x,y$. Accordingly, the terms in \eqref{second renyi entropy final exact} corresponding to $\gamma=x,y$ were dropped. In producing the data for the all-spin-up quench protocol in figures \ref{fig:MutualInformation} and \ref{fig:FlipAndNoFlip} we have used the same $\mathcal{S}^z$ as for the flipped-spin protocol, while the terms for $\gamma=x,y$ do not contribute and have been dropped (apart from the exception $|A|=|B|=|C|=3$ for which we have defined the strings properly, but this particular point is not important anyway).

\section{Correlation matrix with strings}\label{appendix correlation matrix with strings}
 
Here we derive the expression \eqref{eq:correlation matrix string workable expression} for the correlation matrix, that is suitable for numerical implementations. We suppose that the string is given by $\mathcal{S}^\gamma=a_{i_1}a_{i_2}\ldots a_{i_m}$ for some positive even integer $m$. We use Wick theorem (see e.g. appendix \ref{appendix wick theorem}) to write the elements in terms of pfaffians,
\begin{equation}
\left(\Gamma_\gamma\right)_{j,\ell}=\delta_{j,\ell}-\mathrm{pf}\begin{pmatrix}
        M & Q \\ -Q^{\mathrm{T}} & N
    \end{pmatrix} / \mathrm{pf}(N) \; ,
\end{equation}
where $M$ is a $2\times 2$ antisymmetric matrix 
\begin{equation}
    M=\begin{pmatrix}
        0 & \braket{a_{f(j)}a_{f(\ell)}} \\ - \braket{a_{f(j)}a_{f(\ell)}} & 0
    \end{pmatrix},
\end{equation}
$N$ is defined in \eqref{matrix N definition string} and $Q$ is a $2\times m$ matrix
\begin{equation}
    Q=\begin{pmatrix}
        \braket{a_{f(j)} a_{i_1}} & \braket{a_{f(j)} a_{i_2}} & \ldots & \braket{a_{f(j)} a_{i_m}}\\
        \braket{a_{f(\ell)} a_{i_1}} & \braket{a_{f(\ell)} a_{i_2}} & \ldots & \braket{a_{f(\ell)} a_{i_m}}
    \end{pmatrix} \; .
\end{equation}
Using the pfaffian identity (see e.g. \cite{GonzalezBallestero2011})
\begin{equation}
    \mathrm{pf}\begin{pmatrix}
        M & Q \\ -Q^{\mathrm{T}} & N
    \end{pmatrix} = \mathrm{pf}(N)\mathrm{pf} \left(M+QN^{-1}Q^T\right)
\end{equation}
and the property $\mathrm{pf} \left(M+QN^{-1}Q^T\right)= \left(M+QN^{-1}Q^T\right)_{1,2}$, that is just the definition of the pfaffian for a $2\times 2$ matrix, we get \eqref{eq:correlation matrix string workable expression}.

\section{Wick theorem for fermionic Gaussians constructed with complex antisymmetric matrices}\label{appendix wick theorem}
Here we state and prove the Wick theorem, for Gaussians constructed with complex antisymmetric matrices, directly in the formalism with Majorana fermions. The proof takes ingredients from ref.~\cite{Gaudin1960,PerezMartin2007}. 

We consider a $2^d$ dimensional Fock space associated with Majorana fermions $a_1,a_2,\ldots, a_{2d}$, that satisfy the algebra $\{a_\ell,a_n\}=2\delta_{\ell,n}$. We consider the Gaussian state \eqref{gaussian} and we are interested in the ``expectation values" $\tr[\rho(\Gamma) a_{i_1} a_{i_2}\ldots a_{i_{n}}]$, where $i_1,i_2,\ldots,i_n\in\{1,2,\ldots,2d\}$ are not necessarily distinct indices. For odd $n$ the expectation value is zero. For even $n$, the Wick theorem, given in the following, expresses the expectation value as a sum over all contractions, conveniently expressed as a pfaffian. In the proof we will use the relations \eqref{correlation matrix Gamma W relation} (see \cite{Fagotti2010disjoint}) and \eqref{normalization} (see \cite{Fagotti2010disjoint,Klich2014}), proven independently of the Wick theorem given here. 

\paragraph*{Wick theorem.}
For a Gaussian state $ \rho=\exp(\frac{1}{4}\vec{ a}^\dagger W \vec{ a})/\mathcal{Z}$, with $W$ a complex antisymmetric matrix (not necessarily Hermitian) such that $\mathcal{Z}=\tr \exp(\frac{1}{4}\vec{ a}^\dagger W \vec{ a})\neq 0$, and Majorana fermions $ a_{i_1}, a_{i_2},\ldots,  a_{i_{n}}$ (not necessarily distinct),  with $n$ an even number, we have
\begin{equation}
\braket{ a_{i_1}  a_{i_2}\ldots  a_{i_{n}}}=\pf
\begin{pmatrix}
    0 & \braket{ a_{i_1} a_{i_2}} & \braket{ a_{i_1} a_{i_3}}& \ldots &\braket{ a_{i_1} a_{i_n}} \\
    - \braket{ a_{i_1} a_{i_2}} & 0 & \braket{ a_{i_2} a_{i_3}} & \ldots & \braket{ a_{i_2} a_{i_n}} \\
    -\braket{ a_{i_1} a_{i_3}} & -\braket{ a_{i_2} a_{i_3}} & 0 & \ldots & \braket{ a_{i_3} a_{i_n}}\\
    \vdots & \vdots & \vdots & \ddots & \vdots \\
 -\braket{ a_{i_1} a_{i_n}} &  -\braket{ a_{i_2} a_{i_n}} &  -\braket{ a_{i_3} a_{i_n}} & \ldots & 0
    
\end{pmatrix} \; ,
\end{equation}
where $\braket{ O}$ stands for $\tr( O \rho)$.

\paragraph*{Proof.}
For some operators $ O_1,  O_2,\ldots,  O_{n}$ whose anticommutators are $c$-numbers we have
\begin{equation}\label{eq:step Wick theorem proof}
\begin{split}
        &\tr[ O_1  O_2\ldots  O_{n} \rho]=\sum_{k=2}^{n}(-1)^k\{ O_1, O_k\}\tr[( O_2  O_3\ldots  O_{k-1})( O_{k+1} O_{k+2}\ldots  O_{n}) \rho]\\&-\tr[ O_2  O_3\ldots  O_{n} O_1 \rho],
\end{split}
\end{equation}
that is obtained by commuting $ O_1$ with $ O_{k}$ for $k=2,3,\ldots, n$ and using the property that each anticommutator is a $c$-number to bring it outside the trace. Here we abuse the notation slightly, as we replace the operators proportional to the identity by numbers. Relation \eqref{eq:step Wick theorem proof} holds, in particular, for $ O_k= a_{i_k}$, $k=2,3,\ldots,n$ and $ O_1= a_i$ for any $i\in\{1,2,\ldots, 2d\}$. Now the idea is to commute $ O_1$ with $ \rho$ in the last term in \eqref{eq:step Wick theorem proof}, use the cyclic property of the trace and move the term to the left hand side of the equation.

We use the standard nested commutators identity for complex square matrices (see e.g. Proposition 3.35 in \cite{Hall2015}) 
\begin{equation}\label{eq:identity exponentials nested commutators}
    e^A B e^{-A}=B+[A,B]+\frac{1}{2}[A,[A,B]]+\ldots + \frac{1}{n!}[A,[A,\ldots[A,B]\ldots]]+\ldots,
\end{equation}
where in the $n$-th term $n$ commutators appear. It is easy to show
\begin{equation}
\left[\frac{\vec{ a}^\dagger W \vec{ a}}{4}, a_i\right] =-\sum_{j=1}^{2d} W_{ij} a_j,
\end{equation}
from which we obtain recursively for $n$ nested commutators
\begin{equation}\label{step n commutators}
    \left[\frac{\vec{ a}^\dagger W \vec{ a}}{4},\left[\frac{\vec{ a}^\dagger W \vec{ a}}{4},\ldots\left[\frac{\vec{ a}^\dagger W \vec{ a}}{4}, a_i\right]\ldots\right]\right]=(-1)^n \sum_{j=1}^d \left(W^n\right)_{ij} a_j.
\end{equation}
From identity \eqref{eq:identity exponentials nested commutators} we now get
\begin{equation}\label{eq:step Wick theorem commuting density matrix and Majoranas}
     a_i  \rho = \rho \sum_{j=1}^{2d} \left(e^W\right)_{ij} a_j.
\end{equation}

Using \eqref{eq:step Wick theorem commuting density matrix and Majoranas} and the cyclic property of the trace, from \eqref{eq:step Wick theorem proof} we get
\begin{equation}\label{eq:step Wick theorem proof before inverting}
   \sum_{j=1}^{2d} \left(1+e^W\right)_{ij} \tr[ a_j  a_{i_{2}} a_{i_{3}}\ldots  a_{i_{n}}\rho]=\sum_{k=2}^n(-1)^k\{ a_i, a_{i_k}\}\tr[ a_{i_2} a_{i_3}\ldots  a_{i_{k-1}} a_{i_{k+1}}\ldots  a_{i_n} \rho]
\end{equation}
for $j=1,2,\ldots,2d$. From the assumption of the theorem we have $\mathcal{Z}\neq 0$, and since $\mathcal{Z}^2=\det(e^{\frac{W}{2}}+e^{-\frac{W}{2}})$ it follows that $1+e^W$ is invertible. We can thus multiply \eqref{eq:step Wick theorem proof before inverting} by $(1+e^W)^{-1}_{\ell i}$ and sum over $i$ to get
\begin{equation}
    \tr[ a_\ell  a_{i_{2}}\ldots  a_{i_{n}} \rho]=\sum_{k=2}^n(-1)^k C_{\ell i_k}\tr[ a_{i_2} a_{i_3}\ldots  a_{i_{k-1}} a_{i_{k+1}}\ldots  a_{i_n} \rho],
\end{equation}
where $C_{\ell j}=\sum_{i=1}^{2d}(1+e^W)^{-1}_{\ell i}\{ a_i, a_j\}$. Using Majorana anticommutation relations $\{ a_i,  a_j\}=2\delta_{ij}$ explicitly we get a simplification $C_{\ell j}=2(1+e^W)^{-1}_{\ell j}$. Since the correlation matrix is given by $\Gamma=\tanh(W/2)$ it follows $C_{\ell j}=\braket{ a_\ell  a_j}$. We have thus shown
\begin{equation}
    \tr[ a_{i_1}  a_{i_{2}}\ldots  a_{i_{n}} \rho]=\sum_{k=2}^n(-1)^k \braket{ a_{i_1} a_{i_k}}\tr[ a_{i_2} a_{i_3}\ldots  a_{i_{k-1}} a_{i_{k+1}}\ldots  a_{i_n} \rho] \; .
\end{equation}
The theorem follows by applying the recursive definition of the pfaffian for an $n\times n$ antisymmetric matrix $M$ with $n$ even,
\begin{equation}
    \pf (M)=\sum_{k=2}^n(-1)^k M_{1k} \pf (M_{\hat{1},\hat{k}}),
\end{equation}
where $M_{\hat{1},\hat{k}}$ stands for the matrix with both the first and the $k$-th rows and columns removed.

\bibliography{referencesLast}

\end{document}